\documentclass[twocolumn,showpacs,amsmath,amssymb,prb,floatfix]{revtex4}

\usepackage{graphicx}% Include figure files
\usepackage{dcolumn}% Align table columns on decimal point
\usepackage{bm}% bold math
\usepackage{array}
\usepackage{amsmath,amssymb}

\begin{document}
%\begin{CJK*}{JIS}{} %%%commentout 

\title{Generalized susceptibility of quasi-one dimensional system with 
periodic potential: model for
the organic superconductor (TMTSF)$_2$ClO$_4$ }
\author{
Yasumasa Hasegawa$^1$ 
and 
Keita Kishigi$^2$}
\affiliation{
 $^1$Department of Material Science,
Graduate School of Material Science,
University of Hyogo,
 Hyogo 678-1297, Japan\\
$^2$Faculty of Education, Kumamoto University,
Kurokami 2-40-1, Kumamoto, 860-8555,
Japan
}
\date{\today}
\begin{abstract}
The nesting vector and the magnetic susceptibility 
 of the quasi-one-dimensional system 
having imperfectly nested Fermi surface
are studied analytically and numerically. The magnetic susceptibility 
has the plateau-like maximum in 
%``\textit{the tip of the arrow}'' 
``\textit{sweptback}'' 
region in the
momentum space, which is surrounded by $\mathbf{Q}=(2 k_F, \pi) + \mathbf{q}_i$
($k_F$ is the Fermi wave number, $i=1,3,4$, and  
$\mathbf{q}_1$, %= \left(- \frac{4}{\hbar v_F} t_b',0 \right)$,
$\mathbf{q}_3$ %= \left(  \frac{4}{\hbar v_F} t_b',0 \right)$,
and
$\mathbf{q}_{4}$ are given in this paper). %=\left( \frac{1}{\hbar v_F} 
%\frac{24 t_b'}{\sqrt{1+128 \left(\frac{t_b'}{t_b}\right)^2}+1},
%  \pm 2 \sin^{-1} \left[\frac{8 
%\frac{t_b'}{t_b}}{\sqrt{1+128 \left(\frac{t_b'}{t_b}\right)^2}+1}
%\right] \right)$,  
The best nesting vector, at which
the susceptibility $\chi_0(\mathbf{Q})$ has the absolute maximum at $T=0$,
is obtained near 
but not at the inflection point, $\mathbf{Q}=(2 k_F, \pi)+\mathbf{q}_4$.
The effect of the periodic potential $V$ on the susceptibility
 is studied, which is important for
 the successive transitions of the field-induced spin density wave in
(TMTSF)$_2$ClO$_4$.
We obtain that the sweptback region 
(surrounded by $\mathbf{q}_2$, $\mathbf{q}_3$ and $\mathbf{q}_4$
when $V>0$) 
becomes small as  $V$ increases and
it shrinks to $\mathbf{q}_3$ for $V \geq 4 t_b'$, where $t_b'$ 
gives the degree of imperfect nesting of the Fermi surface,
i.e. the second harmonics of the warping in the Fermi surface.
The occurrence of the sign reversal of the Hall coefficient
in the field-induced spin density wave states is 
discussed to be possible only when 
$V<2 t_b'-2 t_4$, where $t_4$ is the amplitude of the fourth harmonics of the
warping in the Fermi surface.
This gives the novel limitation for the magnitude of $V$. 
\end{abstract}


\maketitle
%%\end{CJK*} %%commentout
%%%%%%%%%%%%%%%%%%%%%%%%%%%%%%%%%%%%%%%%%
\section{Introduction}
%%%%%%%%%%%%%%%%%%%%%%%%%%%%%
\begin{figure}[bt]
\includegraphics[width=0.38\textwidth]{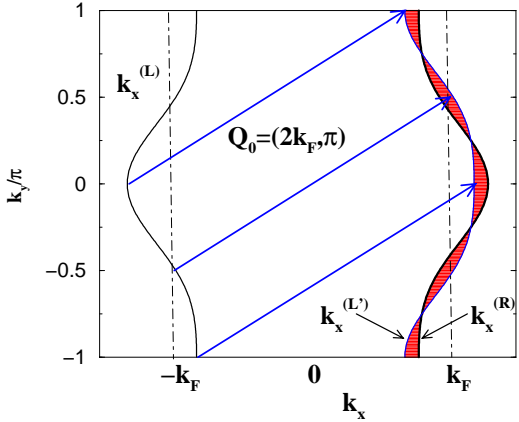}
\caption{
Fermi surface for $V=0$.}
\label{fig1}
\end{figure}
%%%%%%%%%%%%%%%%%%%%%%%%%%%%%%%%%%%%%%%%%%%
%%%%%%%%%%%%%%%%%%%%%%%%%%
\begin{figure}[bt]
\includegraphics[width=0.38\textwidth]{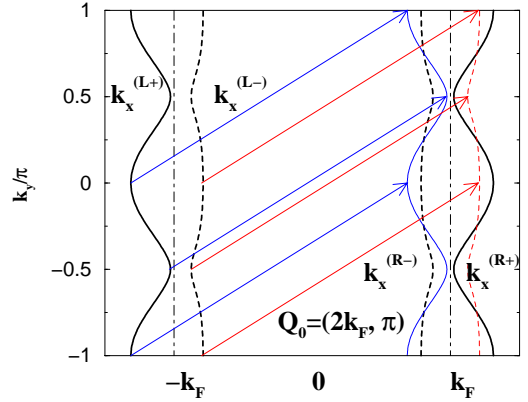}
\caption{
Fermi surface for $V \neq 0$.}
\label{fermisv}
\end{figure}
%%%%%%%%%%%%%%%%%%%%%%%%%%%%%%%%%%%%%%%%%%%
Various interesting properties, such as field-induced spin density wave (FISDW), 
quantum Hall effect and 
superconductivity, have been observed in the quasi-one-dimensional
organic conductors, (TMTSF)$_2$X, 
where X is PF$_6$,  ClO$_4$ etc.\cite{ishiguro98}
The successive transitions between different FISDW phases occur
 as the magnetic field is increased.
The FISDW has been understood as a consequences of the
reduction of the dimensionality due to the magnetic field
and the quantization of the nesting 
vector\cite{gorkov84,montambaux85,yamaji1985,lebed1985,maki1986,virosztek86,chen1987,yamaji1987,machida1993,lebed2002}.
The FISDW phases are characterized by the integer $N$, 
by which the wave number of FISDW is given as
$Q_x=2 k_F + NG$, where $k_F$ is the Fermi wave number, $G=beB/\hbar$, 
$b$ is the lattice constant (we take $b=1$ in this paper), $e$ is the electron
charge,
$B$ is the magnetic field and
$\hbar=h/2\pi$ ($h$ is the Planck constant).
We take $\hbar=1$ hereafter in this paper.
 The Hall conductivity is quantized as $\sigma_{xy}=2 N e^2/h$
with the quantum number $N$ of the nesting 
vector\cite{poilblanc87,yakovenko91,machida94}.
The quantization of the $x$ component of the nesting vector, $Q_x$,
can be seen as  the sharp peaks in the 
susceptibility for the non-interacting system,
$\chi_0(\mathbf{Q})$, at $Q_x=2 k_F + NG$ in the magnetic field.

The peaks of $\chi_0(\mathbf{Q})$ in the magnetic field 
%have
%the origin when the magnetic field is absent. 
can be understood to some extent
by the peaks of $\chi_0(\mathbf{Q})$ in the absence of the magnetic field. 
If the nesting of the Fermi surface is perfect,
$\chi_0(\mathbf{Q})$ 
in the absence of the magnetic
field diverges at the nesting vector as temperature becomes zero.
In that case the successive transitions of FISDW does not happen.
If the nesting of the Fermi surface is not perfect, the 
best nesting vector at $B=0$, which
gives the maximum of $\chi_0(\mathbf{Q})$,  is located 
in the reciprocal space at
\begin{equation}
\mathbf{Q}=\mathbf{Q}_0 + \mathbf{q} ,
\end{equation}
where 
\begin{equation}
  \mathbf{Q}_0 = (2 k_F, \pi),
\end{equation}
and $\mathbf{q} \neq \mathbf{0}$. 
If $q_x >0$, the quantum number $N$ of FISDW is positive. 
If $q_x <0$ at the best nesting vector,
however,  the the negative $N$ is possible 
in some region of the magnetic field\cite{zanchi96}.

Although (TMTSF)$_2$PF$_6$ is well understood 
by the quasi-one-dimensional model, %(Eq.~(\ref{eqq1d})),
(TMTSF)$_2$ClO$_4$ is a little more complicated.
Below $T_{AO} \approx 24$~K
the anion ClO$_4$, which has no inversion symmetry, 
orders alternatively in $y$ direction, 
resulting the
periodic potential $V$ in the electron system.
%%%
Actually, the magnetic field and temperature phase diagram in 
(TMTSF)$_2$PF$_6$\cite{Kwak1981,Chaikin1983,Ribault83,Naughton1985,Kang1993}
 is different from that in 
(TMTSF)$_2$ClO$_4$\cite{MaKernan1995,Scheven1995,Matsunaga2002}. 
%%%
The origin of the different phase diagrams in (TMTSF)$_2$PF$_6$
and (TMTSF)$_2$ClO$_4$ is caused by the periodic potential, $V$.
The magnitude of $V$ is first estimated to be the order of $T_{AO} = 24K$,
i.e. $V \ll t_b$.\cite{Lebed1989,osada}
The suppression of the $N=0$ FISDW state\cite{Lebed1989}
and even-$N$ FISDW states\cite{osada} has been shown by the
perturbation in $V$.
On the other hand,
the magnitude of $V$ has been estimated to be $V = 0.83 t_b$
 from the angle dependence of the magnetoresistance   by Yoshino 
et al.\cite{yoshino1999}.
By treating $V$ not in perturbation, a lot of
interesting features,
such as existence of several nesting 
vectors\cite{miyazaki99,sengupta2002,zanchi2001,haddad2005} and
the phase diagram of the FISDW 
states\cite{hasegawa1998,kishigi2007}, has been obtained.
Recently, Yoshino et al.\cite{yoshino2003} has estimated the value to be 
$V = 0.028 t_a$ ($V = 0.34 t_b$ with their estimation $t_a=12 t_b$).
Lebed et al.\cite{lebed2005,ha2006} have estimated the value as
$V = 0.2 t_b$.
The novel estimation of $V$ is given in this paper from the existence of the
sign reversal of the Hall effect.
%The effect of the periodic potential was studied in the perturbation in
%$V/t_b$ and 
%the suppression of the FISDW with even-$N$ has been 
%explained to be caused by
%the periodic potential $V$.\cite{osada}
%The strength of the anion potential $V$ has been estimated from the 
%angle dependence of the magnetoresistance
%%%by Yoshino et al. as $V/t_b \approx 1.0$\cite{yoshino97},
% as $V/t_b \approx 0.28$\cite{yoshino2003} 
% and $V/t_b 
%\approx 0.2$ \cite{lebed2005,ha2006}.
%Although the value of  $V/t_b$ is now believed to be not large,
%many interesting things have been missed if we study only in 
%the perturbation in 
%$V/t_b$.\cite{hasegawa1998,miyazaki99,sengupta2002,zanchi2001,haddad2005,kishigi2007}

In this paper we study the nesting vector and the susceptibility
in the quasi-one dimensional system with imperfectly nested Fermi surface
in the absence of the magnetic field.
The analytic expression of the susceptibility and the 
nearly flat region in the reciprocal space are given analytically 
for the first time in the simple model with $V=0$. The effect of $V$
on the nesting vector and the susceptibility are studied 
in detail numerically. 
 
%We study a simple model 
%We study the quasi-one dimensional system.

\section{model}
We neglect the small dispersion in $k_z$ direction and study
 the tight binding model
in the square lattice with anisotropic transfer integral elements $t_a \gg t_b$.
We take the lattice constant to be $1$.
In the real system the crystal is triclinic
and we have to consider the multiple-transverse-transfer integrals\cite{yamaji86}  
 but most of the essential features are 
obtained by studying the simple model in the square lattice\cite{ishiguro98}.
 The energy dispersion can be linearized with respect to 
$k_x$ and we take account of the higher harmonic terms for $k_y$ as
\begin{align}
 \epsilon(\mathbf{k}) &= %\hbar 
v_F (|k_x| - k_F) 
 + t_{\perp}(k_y),
\label{eqq1d}
\end{align}
where
\begin{align}
 t_{\perp}(k_y) = &-2 t_b \cos k_y -2 t_b' \cos (2 k_y) 
\nonumber \\
 &-2 t_3 \cos (3 k_y)- 2 t_4 \cos(4 k_y).
\end{align}
and we study the case $t_b$, $t_b'$, $t_3$ and $t_4$ to be positive.
The terms proportional to $t_3$ and $t_4$ are thought to 
be essential\cite{zanchi96,kishigi2007} to 
understand the negative $N$ phase\cite{ribault,matsunaga2} of FISDW in 
some region of the magnetic field.
The Fermi surface consists of two 
``Fermi lines'' near $k_x \approx \pm k_F$, as
shown in Fig.~\ref{fig1}.
The Fermi surface is almost nested, i.e. when we translate 
the left part of the Fermi line  with the vector
$\mathbf{Q} \approx \mathbf{Q}_0$,
it overlaps with the right part of the 
Fermi line, but the overlap is not perfect due to 
the $t_b'$ and $t_4$ terms.

The Brillouin zone is divided into halves in the $k_y$ direction
by the periodic potential.
The Hamiltonian is written as a $2 \times 2$ matrix with 
the anion potential $V$ as
\begin{equation}
 \cal{H}= \left(
\begin{array}{cc}
 \epsilon(\mathbf{k}) & V \\
 V & \epsilon(\mathbf{k}+\mathbf{Q}_A)
\end{array}
\right) ,
\end{equation}
where $\mathbf{Q}_A = (0,\pi)$.
The energy $E(\mathbf{k})$ is given by
\begin{align}
 E(\mathbf{k}) &= \frac{1}{2} 
\biggl( 
 \epsilon(\mathbf{k}) + \epsilon(\mathbf{k}+\mathbf{Q}_A)
\nonumber \\
&\ \ 
\pm \sqrt{\left(\epsilon(\mathbf{k}) 
 - \epsilon(\mathbf{k}+\mathbf{Q}_A)\right)^2+4 V^2 } \  \biggr),
\end{align}
and the Fermi surface consists of four 
lines as shown in Fig.~\ref{fermisv}.

It is known\cite{miyazaki99}  that 
the susceptibility $\chi_0(\mathbf{Q})$ has maximum near $\mathbf{Q} \approx 
\mathbf{Q}_0$ 
 if $V \lesssim 1.5 t_b$ when $t_b'=0.1 t_b$ (i.e. $V \lesssim 15 t_b'$), 
while the absolute maximum of $\chi_0(\mathbf{Q})$
is located near $\mathbf{Q} \approx (2k_F \pm  2 V /v_F, \pi/2)$
if $V \gtrsim 1.5 t_b$.
The peak of  $\chi_0(\mathbf{Q})$ near $\mathbf{Q} \approx 
\mathbf{Q}_0$ is caused by the nesting between  the 
outer Fermi surface 
and the inner Fermi surface
 ($\mathbf{k}_x^{(R+)}$ and $\mathbf{k}_x^{(L-)}$), i.e.,
the red and blue arrows in Fig.~\ref{fermisv},
while
the peaks of  $\chi_0(\mathbf{Q})$ near  
$\mathbf{Q} \approx (2k_F \pm 2V/ %\hbar 
v_F, \pi/2)$
are caused by the nesting the outer Fermi surfaces
($\mathbf{k}_x^{(R+)}$ and $\mathbf{k}_x^{(L+)}$) 
or the inner Fermi surfaces
($\mathbf{k}_x^{(R-)}$ and $\mathbf{k}_x^{(L-)}$)
\cite{kishigi1997,kishigi1998,hasegawa1998}.
The maximum value of $\chi_0(\mathbf{Q})$ 
near $\mathbf{Q} \approx \mathbf{Q}_0$ depends
weakly on $V$ if $V \lesssim 0.4 t_b$, and 
it decreases as $V$ increases if $V \gtrsim 0.4 t_b$.
Sengupta and Dupuis\cite{sengupta2002}
and Zanchi and Bjelis\cite{zanchi2001}  obtained the similar results.

In this paper we examine in detail 
the nesting properties of the quasi-one dimensional systems without and with 
the periodic potential ($V \lesssim 0.5 t_b$).
Thus we focus on the nesting condition for 
only $\mathbf{Q} \approx \mathbf{Q}_0$.

%\[ \mathbf{q}=(\frac{1}{\hbar v_F} 4 t_b',0) \]

\section{nesting of the Fermi surface for $V=0$}
%%%%%%%%%%%%%%%%%%%%%%%%%%%%%
\begin{figure}[bth]
\includegraphics[width=0.38\textwidth]{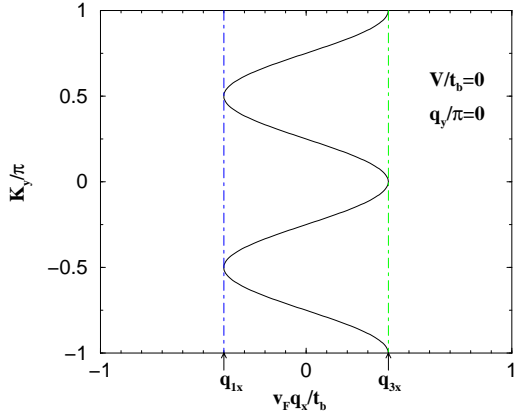}
\caption{
$q_x$ vs $K_y$ (Eq.~\ref{eqqx}) for $q_y=0$ and $q_y=\pi$.}
\label{fig2a}
\end{figure}
%%%%%%%%%%%%%%%%%%%%%%%%%%%%%%%%%%%%%%%%%%%
%%%%%%%%%%%%%%%%%%%%%%%%%%%%%
\begin{figure}[bth]
\includegraphics[width=0.38\textwidth]{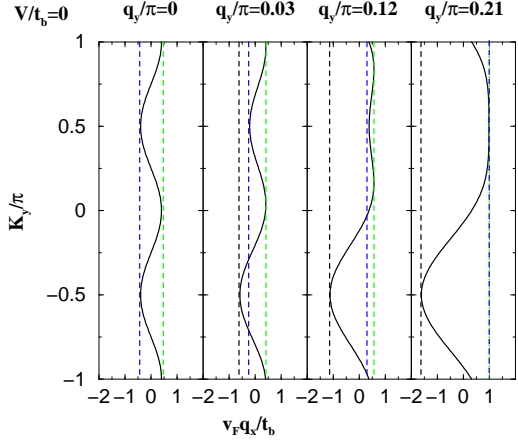}
\caption{$q_x$  vs $K_y$ (Eq.~(\ref{eqqx})) for some values of $q_y$.
There are two minimums ($q_x^{min(\pm)}(q_y)$) and one maximum
 ($q_x^{max}(q_y)$) of $q_x$ as a function of $K_y$ for each
 $0 < |q_y| <q_{4y}$, while only one minimum and one maximum of
$q_x$ for $|q_y| > q_{4y}$ as shown by dotted vertical lines.}
\label{fig2b}
\end{figure}
%%%%%%%%%%%%%%%%%%%%%%%%%%%%%%%%%%%%%%%%%%%
%%%%%%%%%%%%%%%%%%%%%%%%%%%%%
\begin{figure}[tbh]
\includegraphics[width=0.38\textwidth]{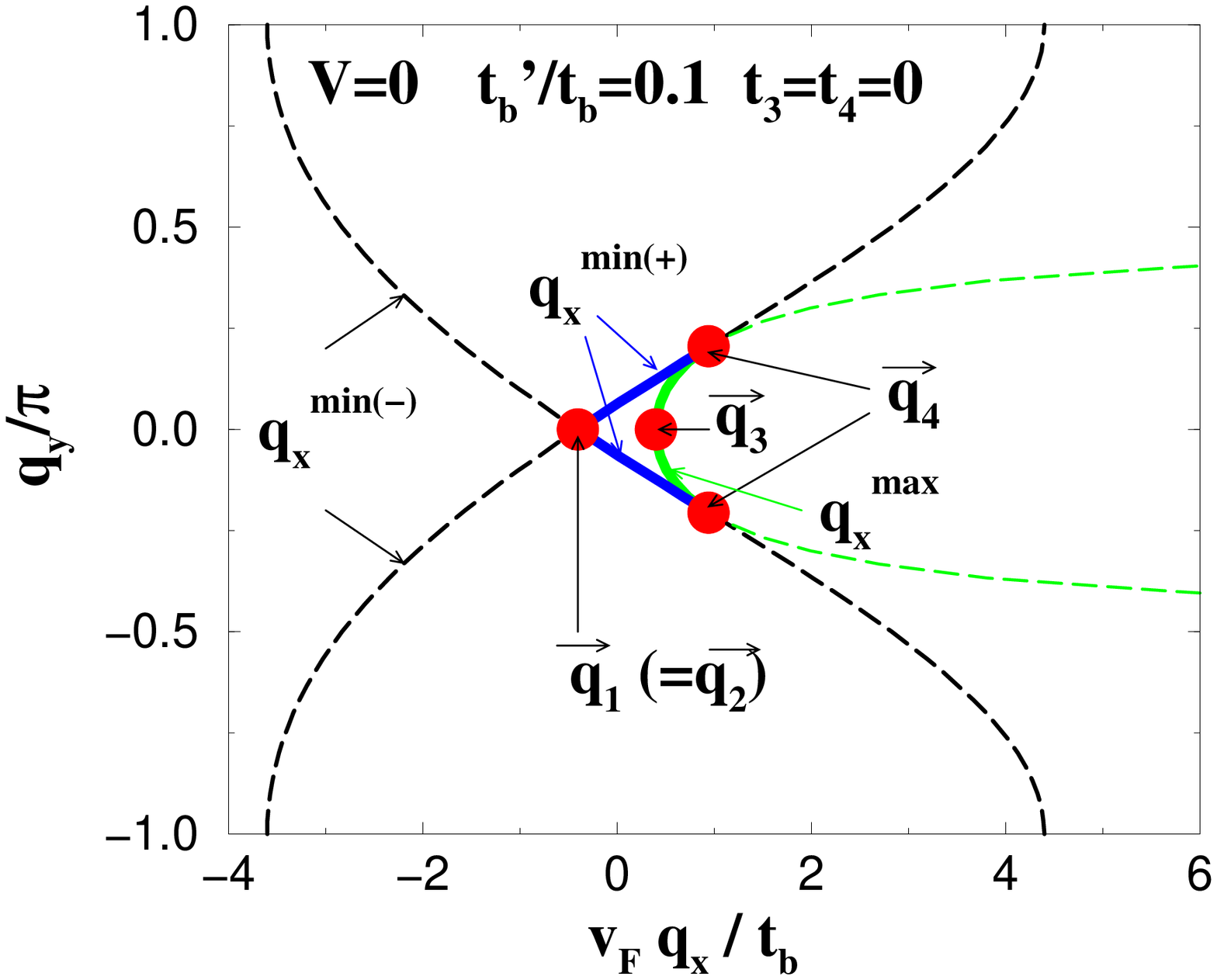}
\includegraphics[width=0.38\textwidth]{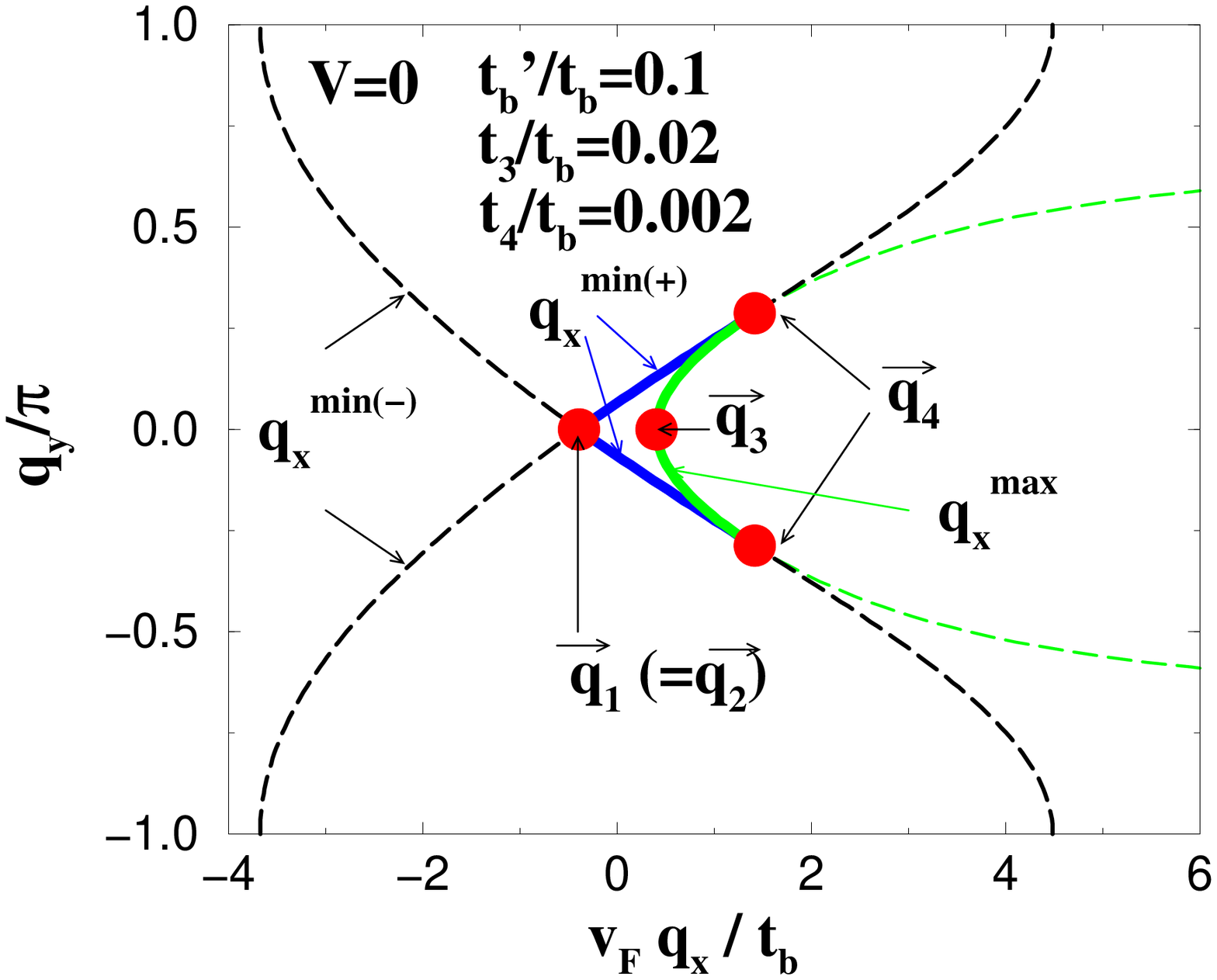}
\caption{
$q_x^{min(-)}(q_y)$ (dashed lines in $q_x < q_{1x}$), 
$q_x^{min(+)} (q_y)$ 
(thick blue lines in $q_{1x} < q_x < q_{4x}$ 
  and dashed lines in $q_x > q_{4x}$) 
and $q_x^{max} (q_y)$
(thick green lines in $q_{3x} < q_x < q_{4x}$ and
  dashed green lines in $q_x > q_{4x}$).
We take $t_b'/t_b=0.1$, $t_3=t_4=0$ (upper figure) and $t_3/t_b=0.02$, 
$t_4/t_b=0.002$ (lower figure). 
In the \textit{sweptback} region 
enclosed by $\mathbf{q}_1$, $\mathbf{q}_3$ and $\mathbf{q}_4$,
$\chi_0(\mathbf{Q})$ has large values.}
\label{figpd}
\end{figure}
%%%%%%%%%%%%%%%%%%%%%%%%%%%%%%%%%%%%%%%%%%%
In this section we study the nesting properties
of the quasi-one dimensional system described by Eq.~(\ref{eqq1d}).
The Fermi surface consists of two curves (see Fig.~\ref{fig1}).
The right and left part of the Fermi surface are given as a function of $k_y$,
\begin{align}
k_x^{(R)}(k_y) &=  k_F - \frac{1}
%{\hbar v_F} t_{\perp}(k_y),
{v_F} t_{\perp}(k_y),
\\
k_x^{(L)}(k_y) &= - k_F +\frac{1}
%{\hbar v_F} t_{\perp}(k_y).
{v_F} t_{\perp}(k_y).
\end{align}
We translate the left part of the Fermi surface with the nesting vector,
$\mathbf{Q} = \mathbf{Q}_0 + \mathbf{q}$.
The translated curve is given by
\begin{align}
 k_x^{(L')}(k_y) 
%&= 2 k_F+k_x^{(L)}(k_y+\pi+q_y)
%\nonumber \\
&=  k_F+ q_x +
%\frac{1}{\hbar v_F} 
\frac{1}{v_F} 
t_{\perp}(k_y +q_y +\pi) .
\end{align}
The difference of the right part of the Fermi surface and the translated 
left part of the Fermi surface is given by
\begin{align}
&k_x^{(L')}(k_y)-k_x^{(R)}(k_y)  \nonumber \\
&= q_x + 
%\frac{1}{\hbar v_F}
\frac{1}{v_F}
 \left(
 t_{\perp}(k_y) + t_{\perp}(k_y+q_y+\pi)\right). 
%%\nonumber \\ 
%%&= q_x -4\frac{1}{\hbar v_F} \huge[
%% t_b \sin (K_y) \sin(\frac{q_y}{2}) 
%%+t_b' \cos (2 K_y) \sin({q_y}) 
%%\nonumber \\
%%&+ t_3 \sin (3 K_y) \sin(\frac{3q_y}{2}) 
%%+t_4 \cos (4 K_y) \sin(2{q_y})\huge] ,  
\end{align}
%%where 
%%\begin{equation}
%% K_y=k_y+\frac{q_y}{2}
%%\end{equation}

If $t_b' = t_4=0$, the nesting of  the Fermi
surface is perfect with $q_x=q_y=0$, i.e.
 $k_x^{(L')}(k_y)-k_x^{(R)}(k_y) =0 $ for all values of 
$k_y$.
If $t_b' \neq 0$ or $t_4 \neq 0$, the nesting of the Fermi surface 
is not perfect. 
In this case
the Fermi surface intersect with the translated one with 
the nesting vector $\mathbf{Q}_0 + \mathbf{q}$,
if %$k_x^{(L')}(k_y)-k_x^{(R)}(k_y) =0$, i.e. 
$q_x$ and $q_y$ satisfy 
\begin{align}
 q_x &= 
%\frac{-1}{\hbar v_F} 
\frac{-1}{v_F} 
\left[ t_{\perp}(k_y) + t_{\perp}(k_y+q_y+\pi) \right]
\nonumber \\
 &=
%\frac{4}{\hbar v_F} 
\frac{4}{v_F} 
\Biggl[
 t_b \sin (K_y) \sin(\frac{q_y}{2}) 
+t_b' \cos (2 K_y) \cos({q_y}) 
\nonumber \\
&+ t_3 \sin (3 K_y) \sin(\frac{3q_y}{2}) 
+t_4 \cos (4 K_y) \cos(2{q_y})\Biggr],
\label{eqqx}
\end{align}
for some value of $K_y$,
where 
\begin{equation}
 K_y=k_y+\frac{q_y}{2} .
\label{eqqx0}
\end{equation}
%%For that $\mathbf{q}$, the susceptibility 
%$\chi(\mathbf{q})$ has large values at 
%%low temperatures $k_B T \ll t_b, t_b'$,
%%where $k_B$ is the Boltzmann constant.
Eq.~(\ref{eqqx}) is the condition for the nesting vector 
($\mathbf{Q} = \mathbf{Q}_0+\mathbf{q}$) to realize the intersection
of the translated
 left part of the Fermi surface with the right part of the 
Fermi surface at $k_y$.
In Fig.~\ref{fig2a} we plot $q_x$ vs $K_y$ for  $q_y=0$. 
%In this figure we also plot the curves for $q_y=\pi$,
%which is useeful for the case of $V \neq 0$.
We define two vectors, $\mathbf{q}_1$ and $\mathbf{q}_3$, as 
$q_{1y}=q_{3y}=0$ and 
$q_{1x}$ and $q_{3x}$ being the minimum 
and the maximum of $q_x$ as a function of $K_y$ at $q_y=0$, 
respectively.
When $t_4 \leq t_b'/4$ (in this paper we study only in this case), 
the maximum of $q_x$ as a function of $K_y$
for $q_y=0$ is given at $K_y=0$ and $\pm \pi$,
and the minimum of  $q_x$ as a function of $K_y$
for $q_y=0$ is given at $K_y=\pm \pi/2$, as shown in Fig.~\ref{fig2a};
\begin{align}
\mathbf{q}_1 &= \left( 
%\frac{4}{\hbar v_F} 
\frac{4}{v_F} 
   \left( -t_b'+t_4\right), 0  \right) ,\\
\mathbf{q}_3 &= \left( 
%\frac{4}{\hbar v_F} 
\frac{4}{v_F} 
   \left(  t_b'+t_4\right), 0  \right) .
\end{align}
We define
$\mathbf{q}_2=\mathbf{q}_1$ for $V=0$ and 
we will define $\mathbf{q}_{2}$ for $V \ne 0$ in section \ref{secvne0}.

We plot $q_x$  vs. $K_y$ (Eq.~(\ref{eqqx}))
for some values of $q_y$ in Fig.~\ref{fig2b}.
As seen in Fig.~\ref{fig2b},  $q_x$  as a function of $K_y$ has two minimums 
at $K_y = \pm \pi/2$
($q_{x}^{min(\pm)}(q_y)$)
and one maximum 
at $0 \leq K_y \leq \pi/2$ ($q_{x}^{max}(q_y)$), if
$0< |q_y| < q_{4y}$  ($\mathbf{q}_{4}$ will be given later).
There are one minimum  at $K_y=-\pi/2$ and one maximum at
$K_y=\pi/2$ if $ |q_y| > q_{4y}$.
%We define $q_{x}^{min(\pm)}(q_y)$ and $q_x^{max}(q_y)$ as a function 
%of $q_y$, and $\mathbf{q}_{4}=(q_{4x},q_{4y})$
%as the points where $q_x^{min+}=q_x^{max} (=q_{x4})$ (See Fig.~\ref{figpd}). 
We obtain $q_x^{min(+)}(q_y)$ and $q_x^{min(+)}(q_y)$ as
\begin{align}
 q_x^{min(+)} (q_y)=& 
%\frac{4}{\hbar v_F}
\frac{4}{v_F}
\Biggl( - t_b' \cos q_y + t_b \sin \frac{|q_y|}{2} 
\nonumber \\
& - t_3 \sin \frac{3 |q_y|}{2} 
+ t_4 \cos 2 q_y \Biggr), \\
 q_x^{min(-)} (q_y)=& 
%\frac{4}{\hbar v_F}
\frac{4}{v_F}
\Biggl( - t_b' \cos q_y - t_b \sin \frac{|q_y|}{2} 
\nonumber \\
&+ t_3 \sin \frac{3 |q_y|}{2} 
+ t_4 \cos 2 q_y \Biggr). 
\end{align}

If $t_3$ and $t_4$ are finite, we have to solve the fourth-degree equation
to obtain the expression of $q_x^{max}(q_y)$, but it is easy to 
obtain $q_x^{max}(q_y)$ numerically.
We define $\mathbf{q}_4 = (q_{4x}, q_{4y})$ by
the equation
\begin{equation}
 q_x^{min(+)} (q_{4y}) = q_x^{max}(q_{4y}) = q_{4x}.
\end{equation}
If $t_3=t_4=0$, the simple expressions of 
$q_x^{max}(q_y)$  and $\mathbf{q}_4$
are obtained as
\begin{equation}
 q_x^{max}(q_y) = 
%\frac{4}{\hbar v_F} 
\frac{4}{v_F} 
(t_b' \cos q_y + \frac{t_b^2 
\sin^2 \frac{q_y}{2}}{8 t_b' \cos q_y}),
\label{qxmax}
\end{equation}
\begin{equation}
 q_{4x}= 
%\frac{1}{\hbar v_F} 
\frac{1}{v_F} 
\frac{24 t_b'}{\sqrt{1+128 \left(\frac{t_b'}{t_b}\right)^2}+1}, 
\end{equation}
and
\begin{equation}
 q_{4y}= \pm 2 \sin^{-1} \left[\frac{8 
\frac{t_b'}{t_b}}{\sqrt{1+128 \left(\frac{t_b'}{t_b}\right)^2}+1}
\right].
\end{equation}
Note that $q_x^{max}(q_y)$ has the physical meaning only if 
$|q_y| \leq q_{4y}$, since the analytical form
Eq.~(\ref{qxmax}) obtained in the case of
 $t_3=t_4=0$ and the numerically obtained values
at $|q_y| > q_{4y}$ corresponds to the local maximum of $q_x$ as a function of
$\sin(K_y/2)$ at $|\sin(K_y/2)| >1$.
We plot $q_x^{max}(q_y)$, $q_x^{min(+)}(q_y)$ and $\mathbf{q}_i$ 
($i=1,3$, and $4$) in Fig.~\ref{figpd}.
There are large overlap between the Fermi line and the translated one,
if $\mathbf{q}$ is in the ``\textit{sweptback}'' region 
 with the apexes $\mathbf{q}_1$ and $\mathbf{q}_4$ enclosed by the
thick lines in Fig.~\ref{figpd}.
\section{susceptibility in the Q1D system with $V=0$}
\begin{figure}[tbh]
\includegraphics[width=0.35\textwidth]{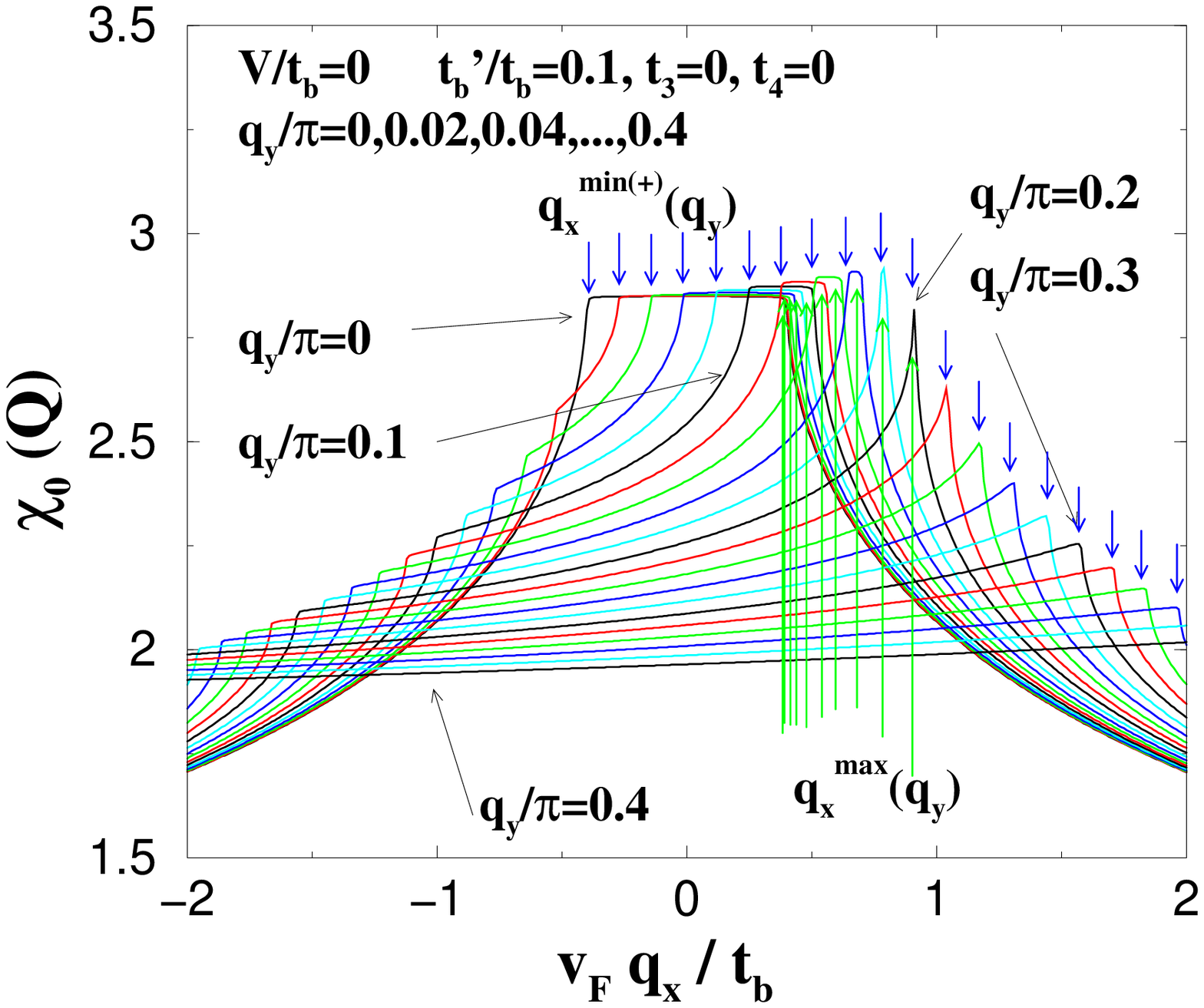}
\includegraphics[width=0.35\textwidth]{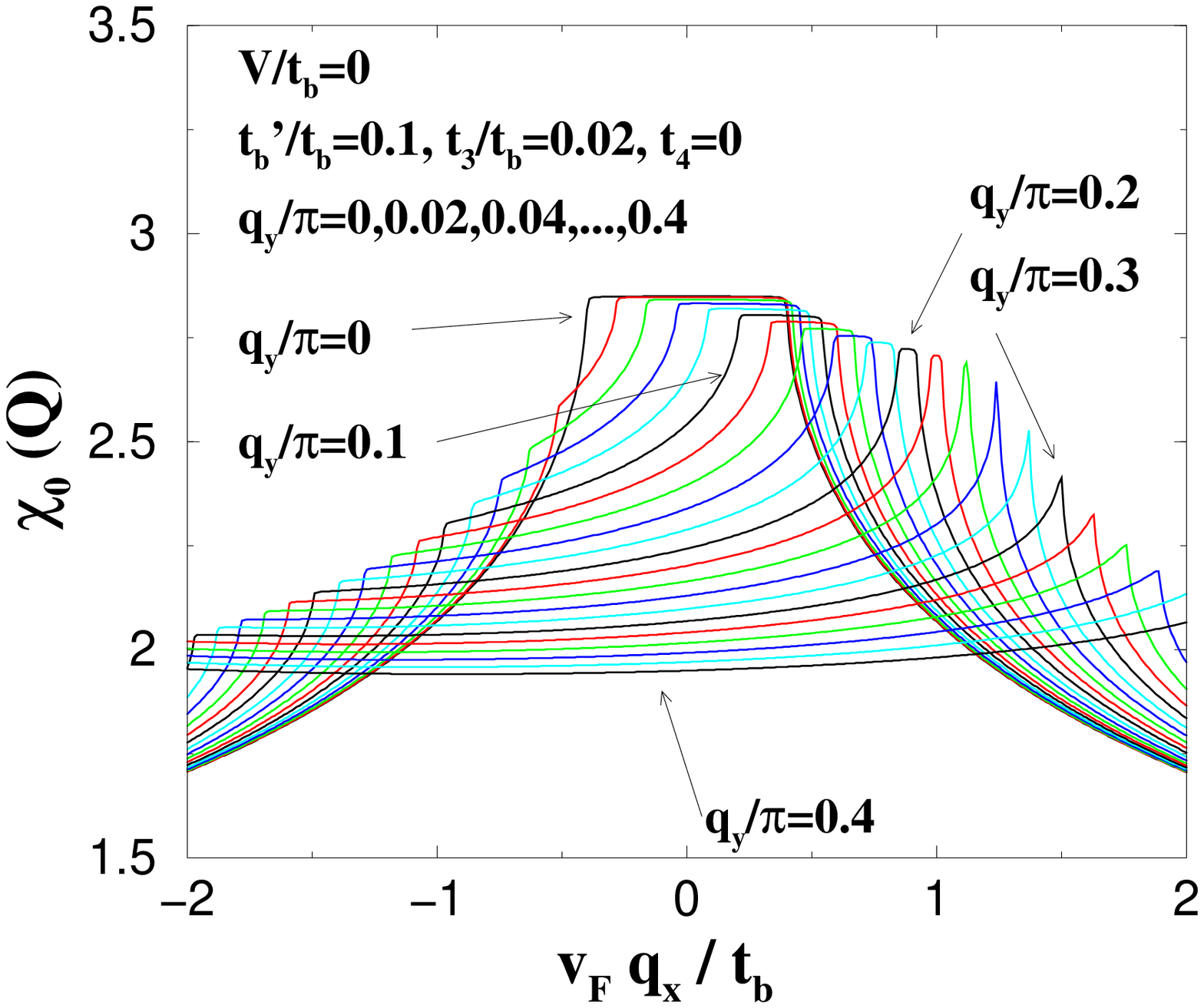}
\includegraphics[width=0.35\textwidth]{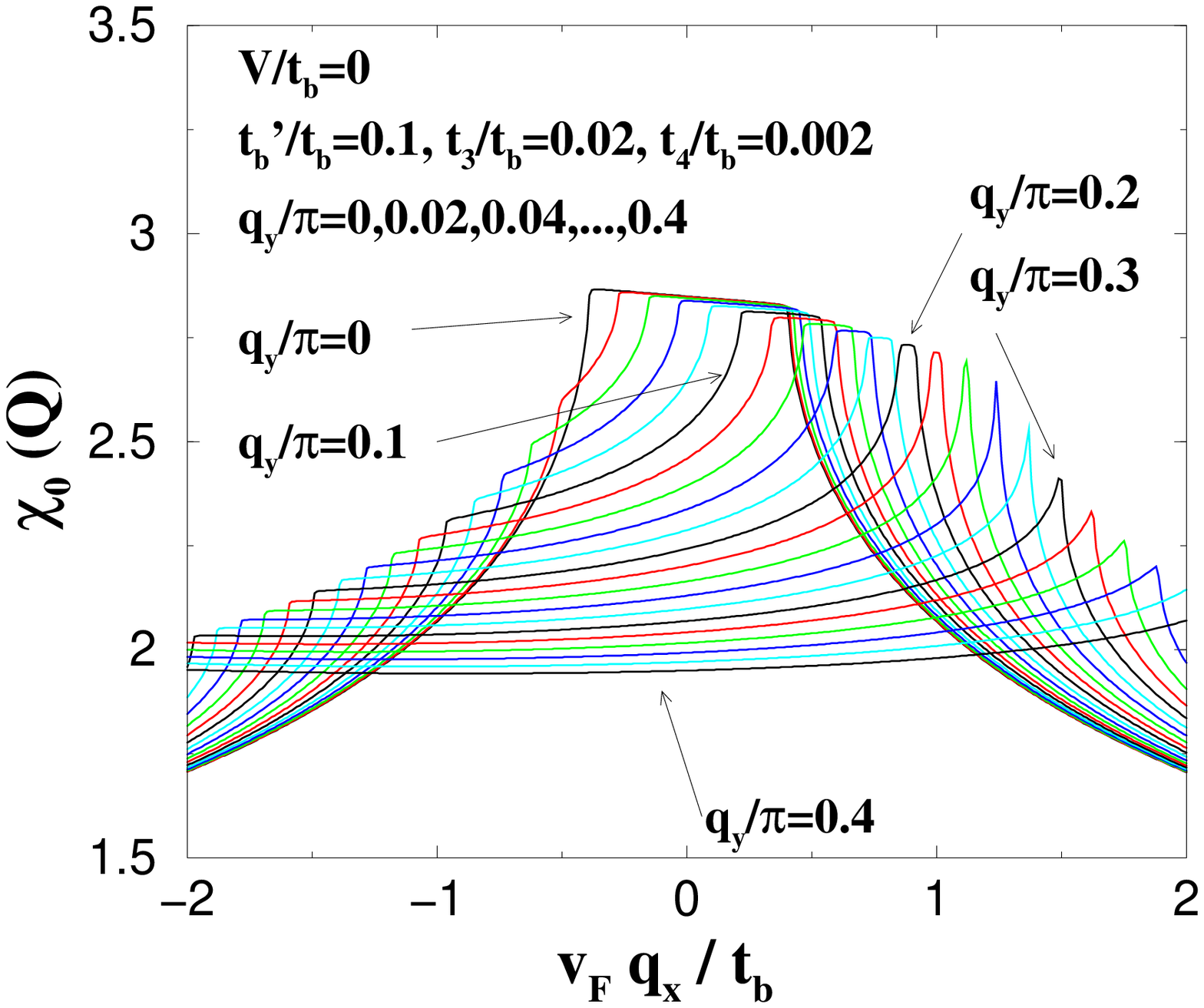}
\caption{
$\chi_0(\mathbf{Q})$ at $T=0$ (eq.~\ref{att0}) 
as a function of $q_x$. We take $t_b'/t_b=0$ and, 
$t_3=t_4=0$ (the upper figure), 
$t_3/t_b=0.02$, $t_4=0$ (middle figure),
and $t_3/t_b=0.02$, $t_4/t_b=0.002$ (lower figure). 
As obtained by Zanchi and Montambaux\cite{zanchi96},
$t_3$ reduces the peak height near $\mathbf{q}_4$ and
$t_4$ lifts the degeneracy at $\mathbf{q}_1$ and $\mathbf{q}_3$.
}
\label{chiqt0}
\end{figure}

The susceptibility
\begin{equation}
\chi_0(\mathbf{Q}) 
 = \sum_{\mathbf{k}}
  \frac{f(E_{\mathbf{k}+\mathbf{Q}})-f(E_{\mathbf{k}})}
   {E_{\mathbf{k}}-E_{\mathbf{k}+\mathbf{Q}}},
\end{equation}
where $f(E_{\mathbf{k}})$ is the Fermi distribution function,
is calculated at $T=0$ as
%We obtain 
%\begin{align}
%\chi(\mathbf{Q}) &= 2 \sum_{\mathbf{k}} \frac{f( \epsilon({\mathbf{k}}))}{\epsilon(\mathbf{k}-\mathbf{Q})-
%\epsilon({\mathbf{k})}}
%\end{align}
%At $T=0$, we obtain
\begin{align}
 & \ \chi_0(\mathbf{Q}) =  \int_{-\pi}^{\pi} \frac{d k_y}{2\pi} 
 \int_{k_x^{(L)}(k_y)}^{k_x^{(R)}(k_y)}  \frac{d k_x}{2\pi} 
 \frac{2} {\epsilon(\mathbf{k}-\mathbf{Q})-
 \epsilon({\mathbf{k})}}
 \nonumber \\
 &= \frac{1}{\pi} \int_{-\pi}^{\pi} \frac{d k_y}{2\pi} \biggl[
%\nonumber \\
%& \ 
 \int_{k_x^{(L)}(k_y)}^{0}  
%%% \frac{d k_x}{\hbar v_F Q_x+t_{\perp}(k_y-Q_y)-t_{\perp}(k_y)}
 \frac{d k_x}{v_F Q_x+t_{\perp}(k_y-Q_y)-t_{\perp}(k_y)}
 \nonumber \\
 &+  \int_{0}^{k_x^{(R)}(k_y)}
%%% \frac{d k_x}{\hbar v_F (-2k_x+Q_x)+t_{\perp}(k_y-Q_y)-t_{\perp}(k_y)} \biggr] 
 \frac{d k_x}{v_F (-2k_x+Q_x)+t_{\perp}(k_y-Q_y)-t_{\perp}(k_y)} \biggr] 
\nonumber \\
&=
%%%\frac{1}{2\pi \hbar v_F}   
\frac{1}{2\pi v_F}   
\int_{-\pi}^{\pi} \frac{d k_y}{2\pi}\biggl[
%%% \frac{\hbar v_F k_F - t_{\perp}(k_y)}
 \frac{v_F k_F - t_{\perp}(k_y)}
%% {\hbar v_F Q_x +t_{\perp}(k_y-Q_y)-t_{\perp}(k_y)}
 {v_F Q_x +t_{\perp}(k_y-Q_y)-t_{\perp}(k_y)}
\nonumber \\
& -\frac{1}{2} \log \left|
 \frac{v_F(Q_x -2 k_F) +  t_{\perp}(k_y-Q_y)+t_{\perp}(k_y)}
 {v_F Q_x + t_{\perp}(k_y-Q_y)-t_{\perp}(k_y)} \right| \biggr]
\label{att0}
\end{align}
The susceptibility is finite at $T=0$ and has the singularity (kinks) as 
a function of $\mathbf{Q}$. 
The singularity of $\chi_0(\mathbf{Q})$ comes from the integration 
of the logarithmic term in eq.~(\ref{att0}).
For $Q_y=\pi$ (i.e. $q_y=0$) and $t_3=t_4=0$, 
%the we obtain 
%\begin{align}
%t_{\perp} (k_y-\pi)+t_{\perp}(k_y) = -4 t' \cos 2 k_y.
%\end{align}
%In this case 
the singular part of $\chi_0(\mathbf{Q}_0+\mathbf{Q})$ is calculated as
\begin{align}
 \chi_{0,sing} &= 
%%%\frac{1}{\pi \hbar v_F} 
\frac{1}{\pi v_F} 
\int_{-\pi}^{\pi} \frac{d k_y}{2\pi} 
\left( -\frac{1}{2} \right)
\log \left|\frac{v_F q_x -4 t_b' \cos 2 k_y}{2 k_F v_F} \right| \nonumber \\
 &= \left\{
 \begin{array}{ll}
%%% -\frac{1}{2 \pi \hbar v_F} 
 -\frac{1}{2 \pi v_F} 
\log \left| 
\frac{q_x v_F + \sqrt{(q_x v_F)^2-(4 t_b')^2}}{4 k_F v_F} \right| &
\\
& \hspace{-1.5cm}
\mbox{if $|q_x v_F| > 4 t_b'$} \\
%%%  -\frac{1}{2 \pi \hbar v_F} 
  -\frac{1}{2 \pi v_F} 
\log \left| \frac{{t_b'}}{k_F v_F} \right| & 
 \\
 & \hspace{-1.5cm} \mbox{if $|q_x v_F| < 4 t_b'$}
\end{array}
\right. .
\label{singular}
\end{align} 
 It is obtained from Eq.~(\ref{singular}), that $\chi_0(\mathbf{q})$ 
has a plateau as a function of $q_x$ 
when $t_3=t_4=0$ and $q_y=0$.
If $t_3$, $t_4$ and $q_y$ are not zero, we have to 
integrate Eq.~(\ref{att0}) numerically.   
In Fig.~\ref{chiqt0} we plot $\chi_0(\mathbf{Q})$ for several 
$t_3$ and $t_4$ and 
 $q_y$ 
as a function of $q_x$. 
It can be seen that if $t_3=t_4=0$, nearly flat peak at 
$q_{x}^{min(+)}(q_y) < q_x < q_x^{max}$ first increases as $q_y$ increases,
and have the absolute maximum before $q_y$ reaches $q_{4y}$ ($=0.2065 \pi$
and $v_F q_{4x}/t_b=0.956$ when $t_b'/t_b=0.1$) as shown in the top figure 
in Fig.~\ref{chiqt0}.
If $t_3 > 0$, the peaks for $q_y\ne 0$ are suppressed as shown in the middle
figure in Fig.~\ref{chiqt0}. If $t_4 >0$, the degeneracy 
of $\chi_0(\mathbf{Q}_0+\mathbf{Q})$
at $\mathbf{q}_1$ and $\mathbf{q}_3$ is lifted and the absolute maximum
of $\chi_0(\mathbf{Q}_0+\mathbf{q})$ is obtained at $\mathbf{q}_1$ for 
the sufficiently large values of $t_3$ and $t_4$,
 as seen in the bottom figure in Fig.~\ref{chiqt0}.
%%%%%%%%%%%%%%%%%%%%%%%%%%%%%
%%%%%%%%%%%%%%%%%%%%%%%%%%%%%%%%%%%%%%%%%%%%%%%%%%%%%%%%%%%%%%%
%%%%%%%%%%%%%%%%%%%%%%%%%%%%%
\begin{figure}[tbh]
\includegraphics[width=0.38\textwidth]{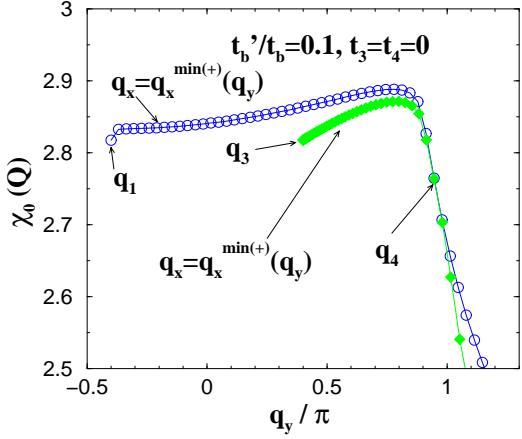}
\caption{
$\chi_0(\mathbf{Q})$ at $T=0$ (eq.~\ref{att0}) 
as a function of $q_x$ for $(q_x, q_y)$ 
on the curves $(q_x^{min(+)}(q_y), q_y)$ 
(filled green diamonds) and $(q_x^{max}(q_y), q_y)$ (open blue circles) 
in Fig.~\ref{figpd}. For $q_x > q_{4x}$ we use Eq.~(\ref{qxmax}), 
although curves $(q_x^{max}, q_y)$
terminate at $\mathbf{q}=\mathbf{q}_4$. 
Note that the absolute maximum is not realized at  $(q_{4x},q_{4y})$.
We take $t_b'/t_b=0.1$, $t_3=t_4=0$.
}
\label{chiqt0x}
\end{figure}
%%%%%%%%%%%%%%%%%%%%%%%%%%%%%%%%%%%%%%%%%%%%%%%%%%%%%%%%%%%%%%%
%%%%%%%%%%%%%%%%%%%%%%%%%%%%%
\begin{figure}[tbh]
\includegraphics[width=0.38\textwidth]{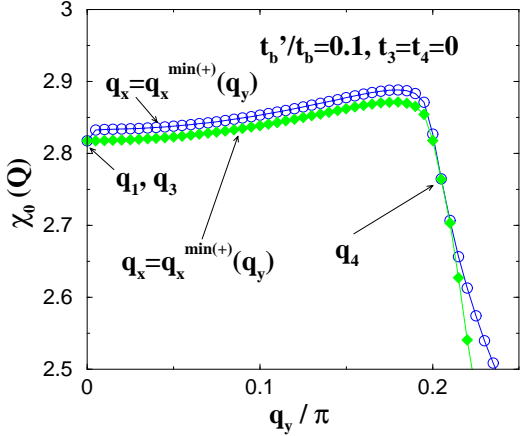}
\caption{
$\chi_0(\mathbf{Q})$ at $T=0$ (eq.~\ref{att0}) 
as a function of $q_y$ for $(q_x, q_y)$ on 
the curves $(q_x^{min(+)}(q_y), q_y)$  (filled green diamonds)
and $(q_x^{max}(q_y), q_y)$ (open blue circles) in Fig.~\ref{figpd}.
 We take $t_b'/t_b=0.1$, $t_3=t_4=0$.}
\label{chiqt0y}
\end{figure}
%\clearpage
%%%%%%%%%%%%%%%%%%%%%%%%%%%%%
%%%%%%%%%%%%%%%%%%%%%%%%%%%%%%%%%%%%%%%%%%%%%%%%%%%%%%%%%%%%%%%%%%%%%%%%%%%%
%\end{document}
\begin{figure}[tb]
\includegraphics[width=0.38\textwidth]{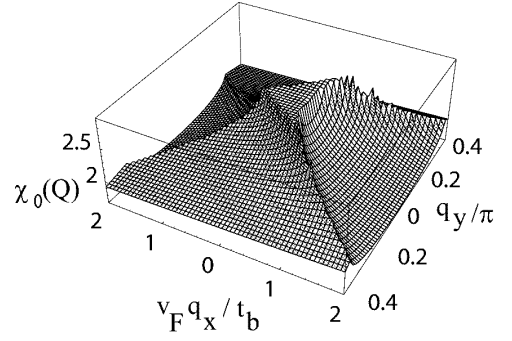}
\caption
{3D plot of
$\chi_0(\mathbf{Q})$ at $T=0$ (eq.~\ref{att0}) 
as a function of $q_x$ and $q_y$. We take $t_b'/t_b=0.1$, $t_3=t_4=0$.}
\label{fig3d}
\end{figure}
%%%%%%%%%%%%%%%%%%%%%%%%%%%%%%%%%%%%%%%%%%%%%%%%

As seen in Fig.~\ref{chiqt0}, 
$\chi_0(\mathbf{Q}_0+\mathbf{q})$ has plateau-like maximum in the region 
$q_x^{min(+)}(q_y) < q_x <q_x^{max}(q_y)$.
The absolute maximum of $\chi_0(\mathbf{Q}_0+\mathbf{q})$ occurs 
at $\mathbf{q}$ close to $\mathbf{q}_4$ 
but not at $\mathbf{q}=\mathbf{q}_4$,
as seen in Figs.~\ref{chiqt0x} and \ref{chiqt0y},
where we plot $\chi_0(\mathbf{Q}_0+\mathbf{q})$ 
as a function of $q_x$ or $q_y$
on the curves of $q_x=q_x^{min(+)}(q_y)$ and $q_x=q_x^{max}(q_y)$,
respectively.
The three-dimensional plot of $\chi_0(\mathbf{Q}_0+\mathbf{q})$ 
is shown in 
Fig.~\ref{fig3d}. % and \ref{contour}.
When $t_3=t_4=0$ and $q_y=q_{4y}$, eq.~(\ref{eqqx}) becomes
\begin{align}
 q_x &= %\frac{1}{\hbar k_F}%
%  \frac{4 t_b'}{\sqrt{1+128 \left(\frac{t_b'}{t_b}\right)^2}+1} 
%  \left(6-4 (Y-1)^2 \right)
%\nonumber \\
%     &=
%%%\frac{1}{\hbar k_F}
\frac{1}{ k_F}
  \frac{4 t_b'
 \left(6-16 \sin^4 (K_y-\frac{\pi}{2}) \right)}
{\sqrt{1+128 \left(\frac{t_b'}{t_b}\right)^2}+1} 
 .
\end{align}
Therefore, $q_x$ as a function of $K_y$ has a maximum at $K_y=\pi/2$ as
$q_x \propto 6-16(K_y-\pi/2)^4$ when $q_y=q_{4y}$. 
With the vector $\mathbf{Q}=\mathbf{Q}_0+\mathbf{q}_4$
the nesting of the Fermi surface 
is better 
than other $\mathbf{q}$'s, 
which will make the expectation 
of the large $\chi_0(\mathbf{Q}_0+\mathbf{q})$.
However, the region of $q_y$ where 
$\chi_0(\mathbf{Q}_0+\mathbf{q})$ is mainly contributed, 
is larger at $q_x \lessapprox 
q_{4x}$ and $|q_y| \lessapprox q_{4y}$ than at  $\mathbf{q}=\mathbf{q}_4$.
This is the reason why the 
absolute maximum of $\chi_0(\mathbf{Q}_0+\mathbf{q})$ is not
located at the inflection point ($\mathbf{q}=\mathbf{q}_4$).
\section{Nesting of the Fermi surface for $V\neq 0$}
\label{secvne0}
In this section we study the effects of periodic potential 
$V$ on the nesting of the Fermi surface and the 
susceptibility.
When $V \neq 0$, there are two pairs of the Fermi lines in
$k_x-k_y$ plane (see Fig.~\ref{fermisv}), 
which are given by $k_x$ as a function of $k_y$,
i.e., $k_x^{L\pm} (k_y)$ and $k_x^{R\pm} (k_y)$
for the left and the right parts of the Fermi lines, respectively.
The nesting vectors are characterized into
four types according to the pairs of the
left and right parts of the Fermi lines, 
i.e. $(+,-)$, $(-,+)$, $(+,+)$, and $(-,-)$ 
as shown in Fig.~\ref{fermisv}.
The left and right parts of the Fermi lines are given by
\begin{align}
  k_x^{(L \pm)} (k_y)&= -k_F - 
%%%\frac{1}{\hbar v_F} 
\frac{1}{v_F} 
\biggl( -t_{\perp}(k_y)-t_{\perp}(k_y+\pi) \nonumber \\
& \pm
\sqrt{[t_{\perp}(k_y)-t_{\perp}(k_y+\pi)]^2 +4 V^2} \biggr) ,
\label{eqkxvl}
\end{align}
and
\begin{align}
 k_x^{(R \pm)} (k_y)&= k_F + 
%%%\frac{1}{\hbar v_F} 
\frac{1}{v_F} 
\biggl( -t_{\perp}(k_y)-t_{\perp}(k_y+\pi) \nonumber \\
&  \pm
\sqrt{[t_{\perp}(k_y)-t_{\perp}(k_y+\pi)]^2 +4 V^2} \biggr) .
\label{eqkxvr}
\end{align}
%%%%%%%%%%%%%%%%%%%%%%%%%%%%%
The condition for the Fermi surface intersect
by the translation of the left part (Eq.~(\ref{eqqx}) for $V=0$)
 is written as the four equations ($++, +-, -+$, and $--$),
\begin{align}
 q_x^{(\pm \pm)} &= 
%%%\frac{1}{2 \hbar v_F} 
\frac{1}{2 v_F} 
\biggl[ -t_{\perp}(k_y)-t_{\perp}(k_y+\pi) \nonumber \\
 &-t_{\perp}(k_y+q_y)- t_{\perp}(k_y+q_y+\pi) \nonumber \\
 &\pm \sqrt{[t_{\perp}(k_y)-t_{\perp}(k_y+\pi)]^2+4 V^2} \nonumber \\
 &\pm \sqrt{[t_{\perp}(k_y+q_y)-t_{\perp}(k_y+q_y+\pi)]^2+4 V^2} \biggr] .
\label{eqqxV}
\end{align}
When $q_y=0$, we obtain Eq.~(\ref{eqqxV}) for   $(+,-)$ and $(-,+)$
to be the same as that for $V=0$ (Eq.~(\ref{eqqx})),
\begin{equation}
 q_x^{(+-)}=q_x^{(-+)}=
%%%\frac{1}{\hbar v_F}
\frac{1}{v_F}
 [-t_{\perp}(k_y)-t_{\perp}(k_y+\pi)] .
\label{eqqxpmq0}
\end{equation}
The condition for the intersect of $(+,+)$ is obtained as 
\begin{align}
 q_x^{(++)}=&
%\frac{1}{\hbar v_F}
\frac{1}{v_F}
 [-t_{\perp}(k_y)-t_{\perp}(k_y+\pi)]
\nonumber \\
%%% &+\frac{1}{\hbar v_F}
 &+\frac{1}{v_F}
 \sqrt{(t_{\perp}(k_y)-t_{\perp}(k_y+\pi))^2+4 V^2} ,
\end{align}
and the condition for the intersect of $(--)$ is obtained as 
\begin{align}
q_x^{(--)}=&
%%%\frac{1}{\hbar v_F}
\frac{1}{v_F}
[-t_{\perp}(k_y)-t_{\perp}(k_y+\pi)]
\nonumber \\
%%%&-\frac{1}{\hbar v_F}
&-\frac{1}{v_F}
\sqrt{(t_{\perp}(k_y)-t_{\perp}(k_y+\pi))^2+4 V^2}.
\end{align}
%%%%%%%%%%%%%%%%%%%%%%%%%%%%%%%
\begin{figure}[tbh]
\includegraphics[width=0.38\textwidth]{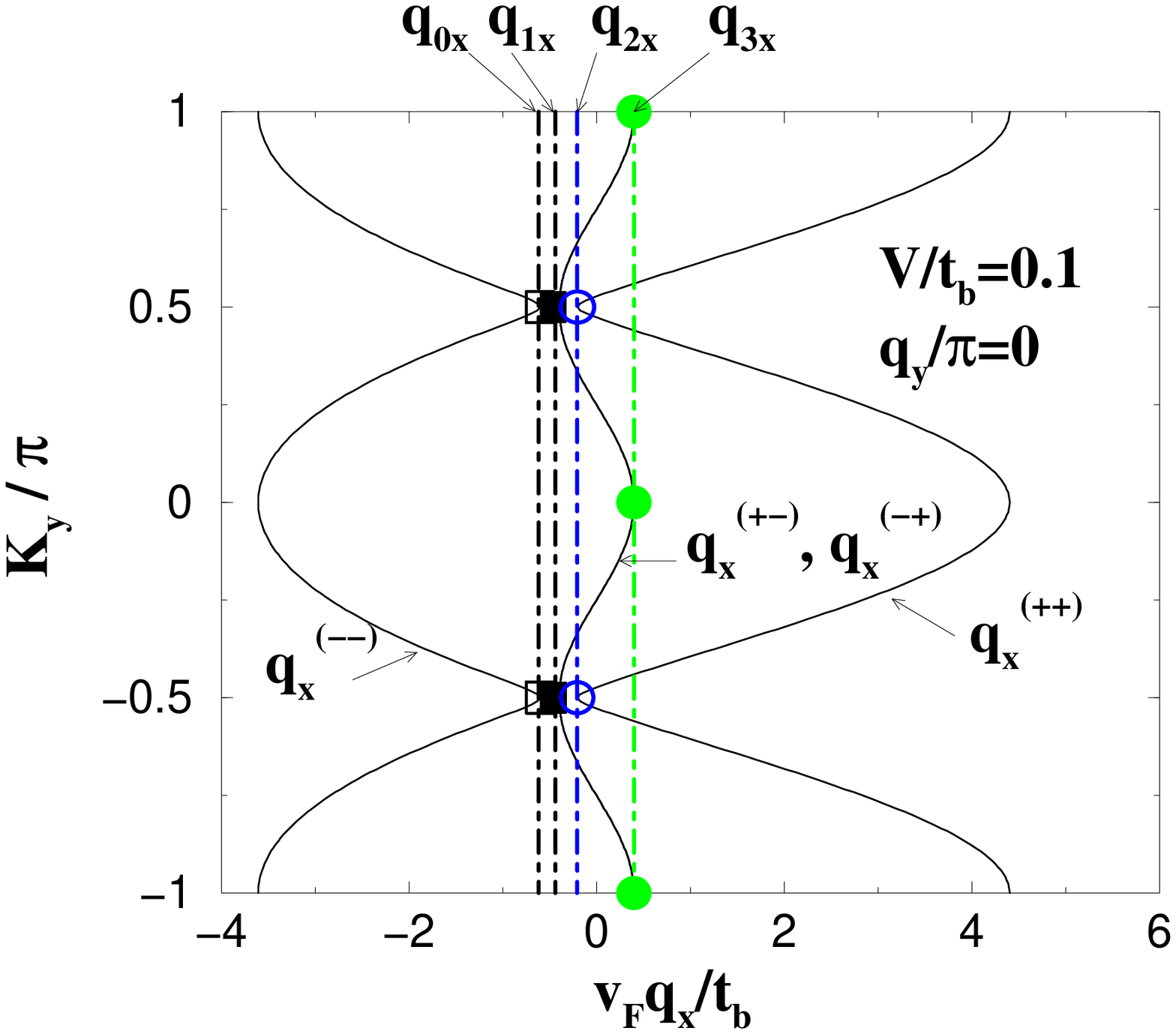}
\caption{
$q_x$ vs. $K_y$ (Eq.~\ref{eqqxV}) for $q_y=0$ and $V/t_b=0.1$.}
\label{figv0}
\end{figure}
%%%%%%%%%%%%%%%%%%%%%%%%%%%%%%%%%%%%%%%%%%%
%%%%%%%%%%%%%%%%%%%%%%%%%%%%%%%
\begin{figure}[tbh]
\includegraphics[width=0.38\textwidth]{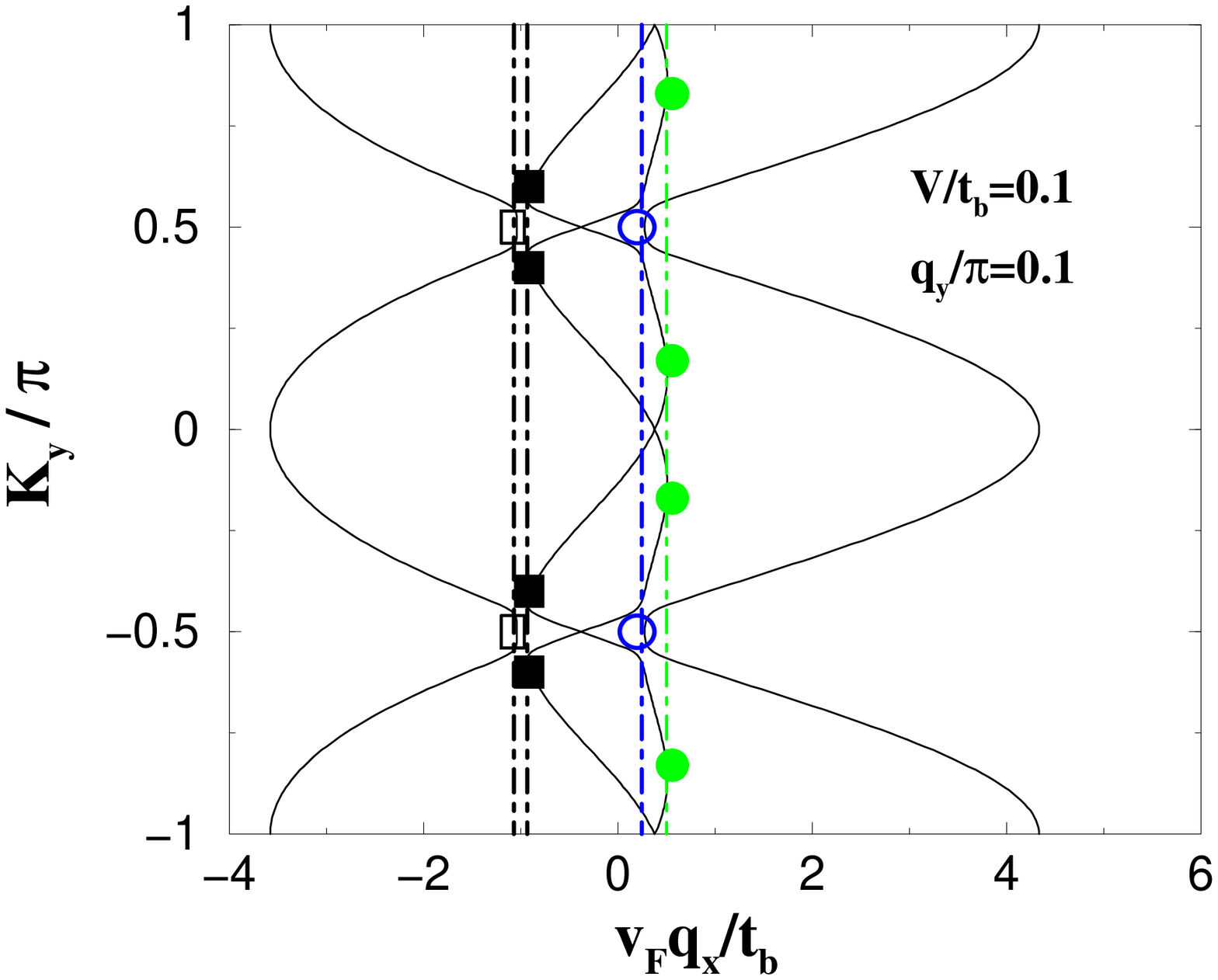}
\caption{
$q_x$ as a function of $K_y$ for $q_y/\pi=0.1$}
\label{figv11}
\end{figure}
%%%%%%%%%%%%%%%%%%%%%%%%%%%%%%%%%%%%%%%%%%%
%%%%%%%%%%%%%%%%%%%%%%%%%%%%%%%
\begin{figure}[tbh]
\includegraphics[width=0.38\textwidth]{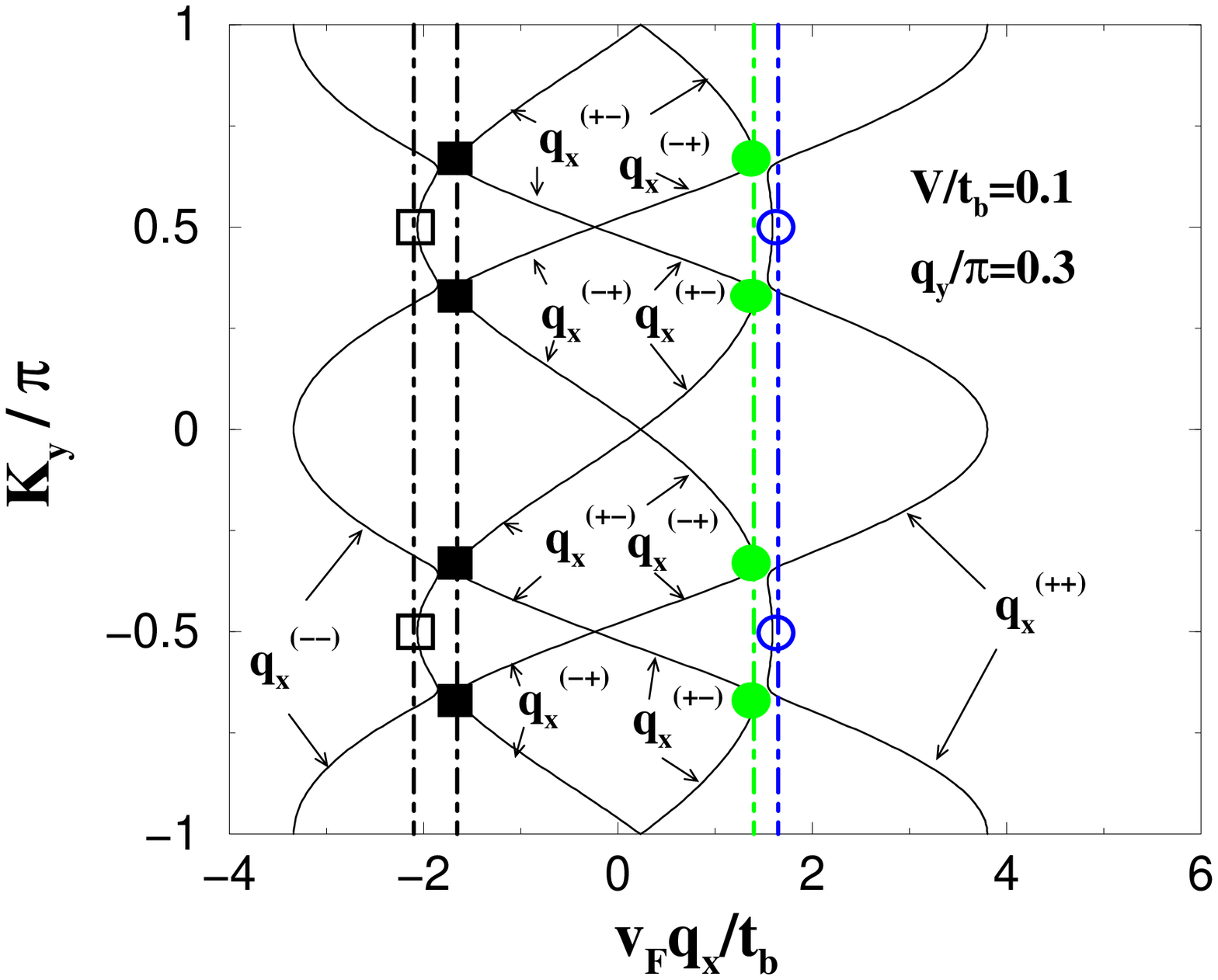}
\caption{
$q_x$ as a function of $K_y$ for $q_y/\pi=0.3$}
\label{figv13}
\end{figure}
%%%%%%%%%%%%%%%%%%%%%%%%%%%%%%%%%%%%%%%%%%%
%%%%%%%%%%%%%%%%%%%%%%%%%%%%%%%
\begin{figure}[tbh]
\includegraphics[width=0.38\textwidth]{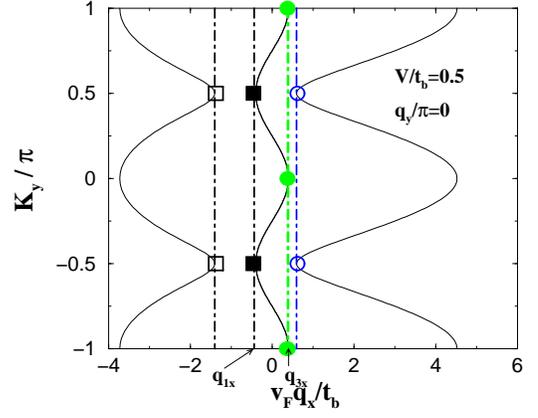}
\caption{
$q_x$ as a function of $K_y$ for $q_y/\pi=0.0$ and $V/t_b=0.5$.}
\label{figv50}
\end{figure}
%%%%%%%%%%%%%%%%%%%%%%%%%%%%%%%%%%%%%%%%%%%
%%%%%%%%%%%%%%%%%%%%%%%%%%%%%%%
\begin{figure}[tbh]
\includegraphics[width=0.38\textwidth]{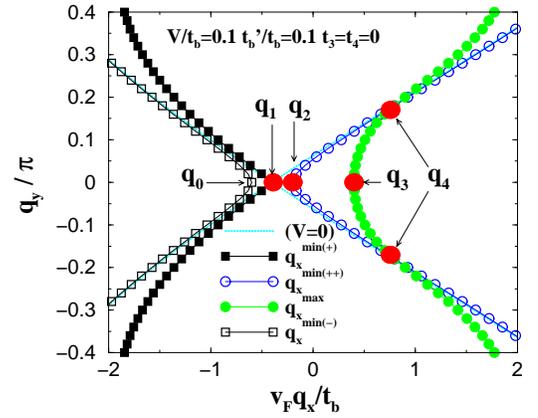}
\caption{
The same as Fig.~\ref{figpd} for $V/t_b =0.1$.
$q_1$ is given as the  minimum of $q_x^{(+-)}$ as a function of $K_y$ 
for each $q_y$.
$q_2$ is given as $q_x^{(++)}$ at $K_y=\pi/2$ for each $q_y$.
$q_3$ is given as the maximum of $q_x^{(+-)}$ as a function of $K_y$ 
for each $q_y$.
$q_0$ is given as $q_x^{(--)}$ at $K_y=\pi/2$ for each $q_y$.
}
\label{regionv1}
\end{figure}
%%%%%%%%%%%%%%%%%%%%%%%%%%%%%%%
%%%%%%%%%%%%%%%%%%%%%%%%%%%%%%%%
%\begin{figure}[tbh]
%\includegraphics[width=0.38\textwidth]{regionv2.eps}
%\caption{
%The same as Fig.~\ref{regionv1} for $V/t_b =0.2$.}
%\label{regionv2}
%\end{figure}
%%%%%%%%%%%%%%%%%%%%%%%%%%%%%%%
%%%%%%%%%%%%%%%%%%%%%%%%%%%%%%%
\begin{figure}[tbh]
\includegraphics[width=0.38\textwidth]{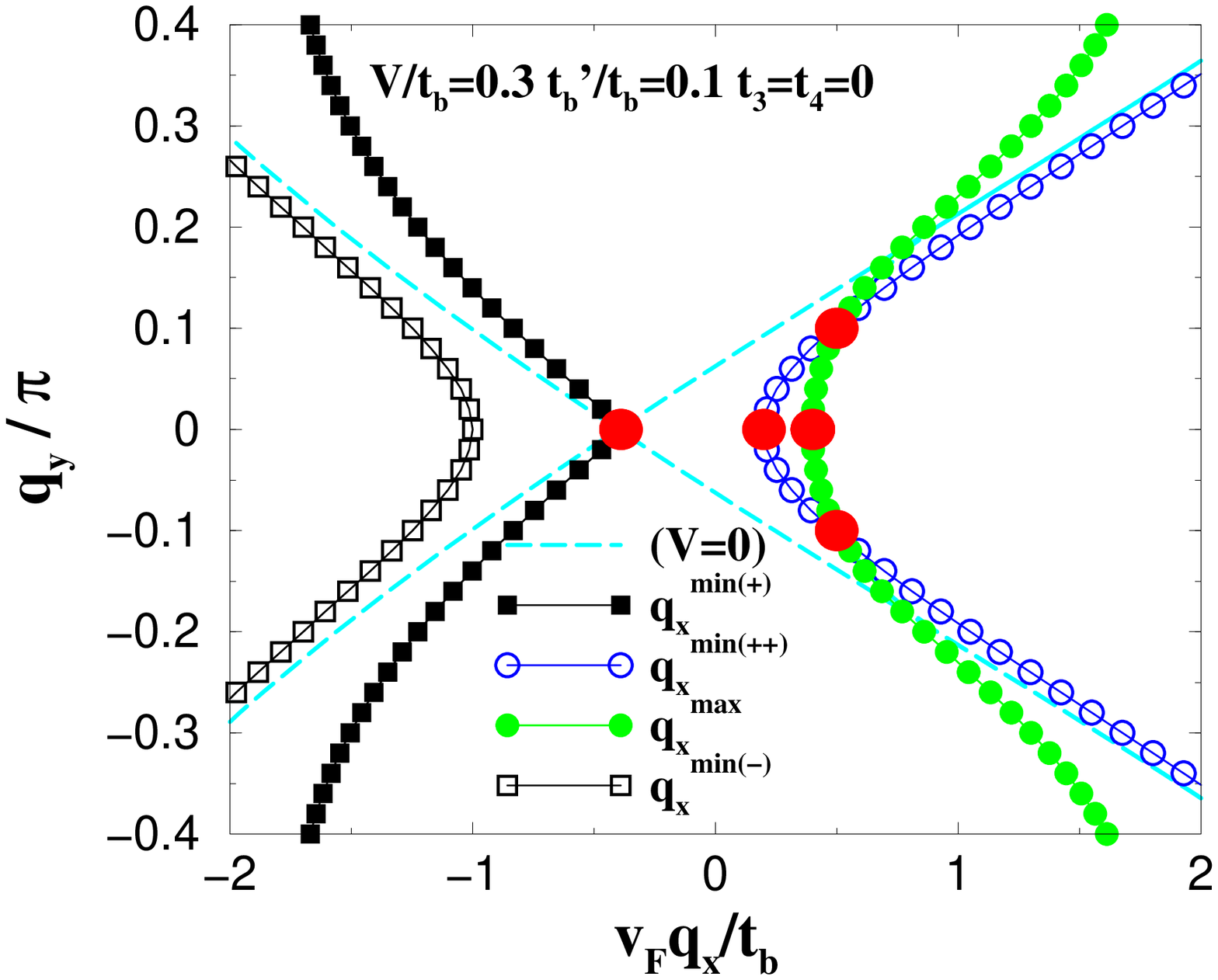}
\caption{
The same as Fig.~\ref{regionv1} for $V/t_b =0.3$.}
\label{regionv3}
\end{figure}
%%%%%%%%%%%%%%%%%%%%%%%%%%%%%%%
%%%%%%%%%%%%%%%%%%%%%%%%%%%%%%%
\begin{figure}[tbh]
\includegraphics[width=0.38\textwidth]{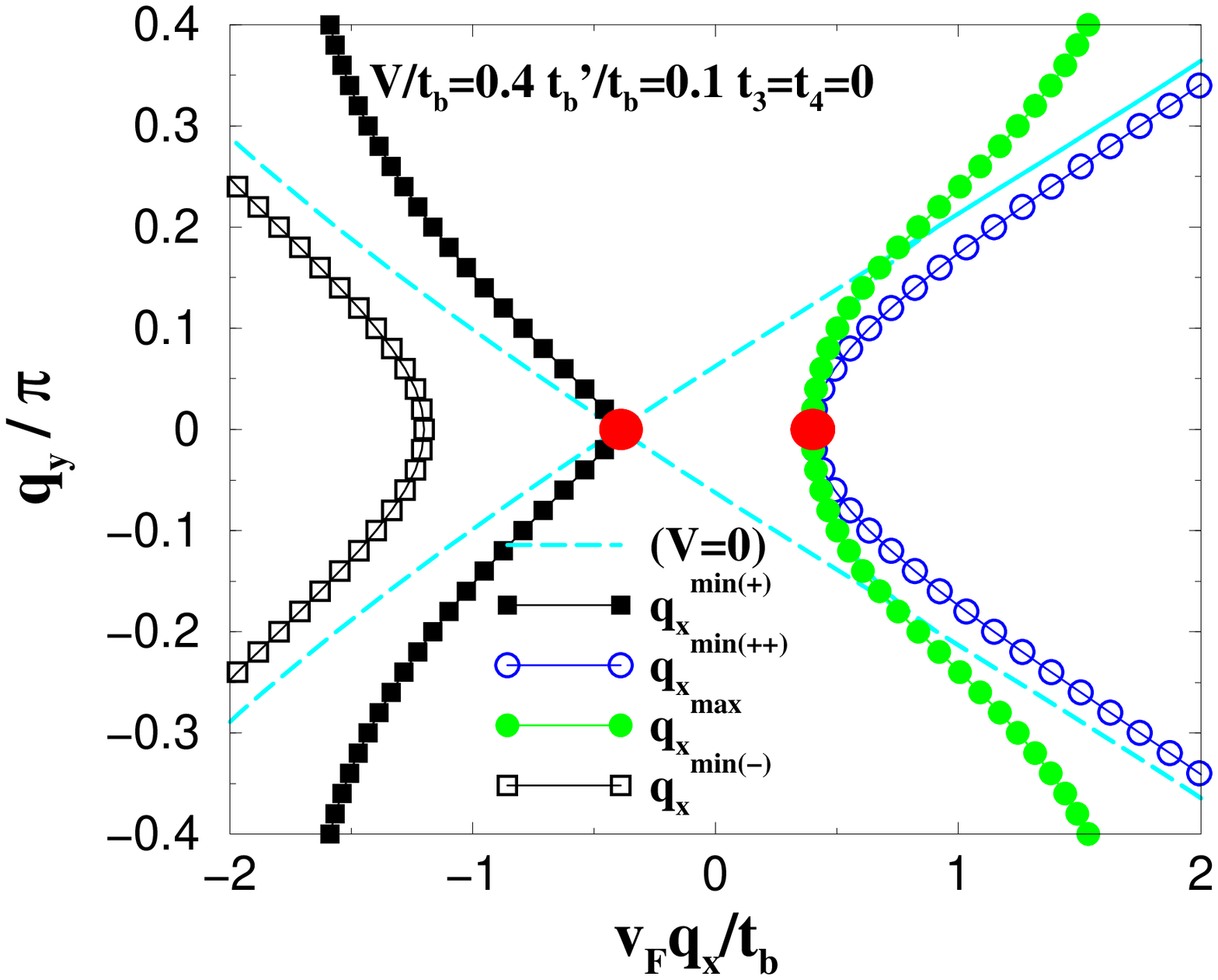}
\caption{
The same as Fig.~\ref{regionv1} for $V/t_b =0.4$.}
\label{regionv4}
\end{figure}
%%%%%%%%%%%%%%%%%%%%%%%%%%%%%%%
%%%%%%%%%%%%%%%%%%%%%%%%%%%%%%%%
%\begin{figure}[tbh]
%%\includegraphics[width=0.38\textwidth]{t3400v5y02.eps}
%\includegraphics[width=0.38\textwidth]{fig18HK.eps}
%\caption{
%$q_x$ as a function of $K_y$ for $q_y/\pi=0.0$ and $V/t_b=0.5$.}
%\label{figv52}
%\end{figure}
%%%%%%%%%%%%%%%%%%%%%%%%%%%%%%%%%%%%%%%%%%%
%
%%%%%%%%%%%%%%%%%%%%%%%%%%%%%%%
\begin{figure}[tbh]
\includegraphics[width=0.38\textwidth]{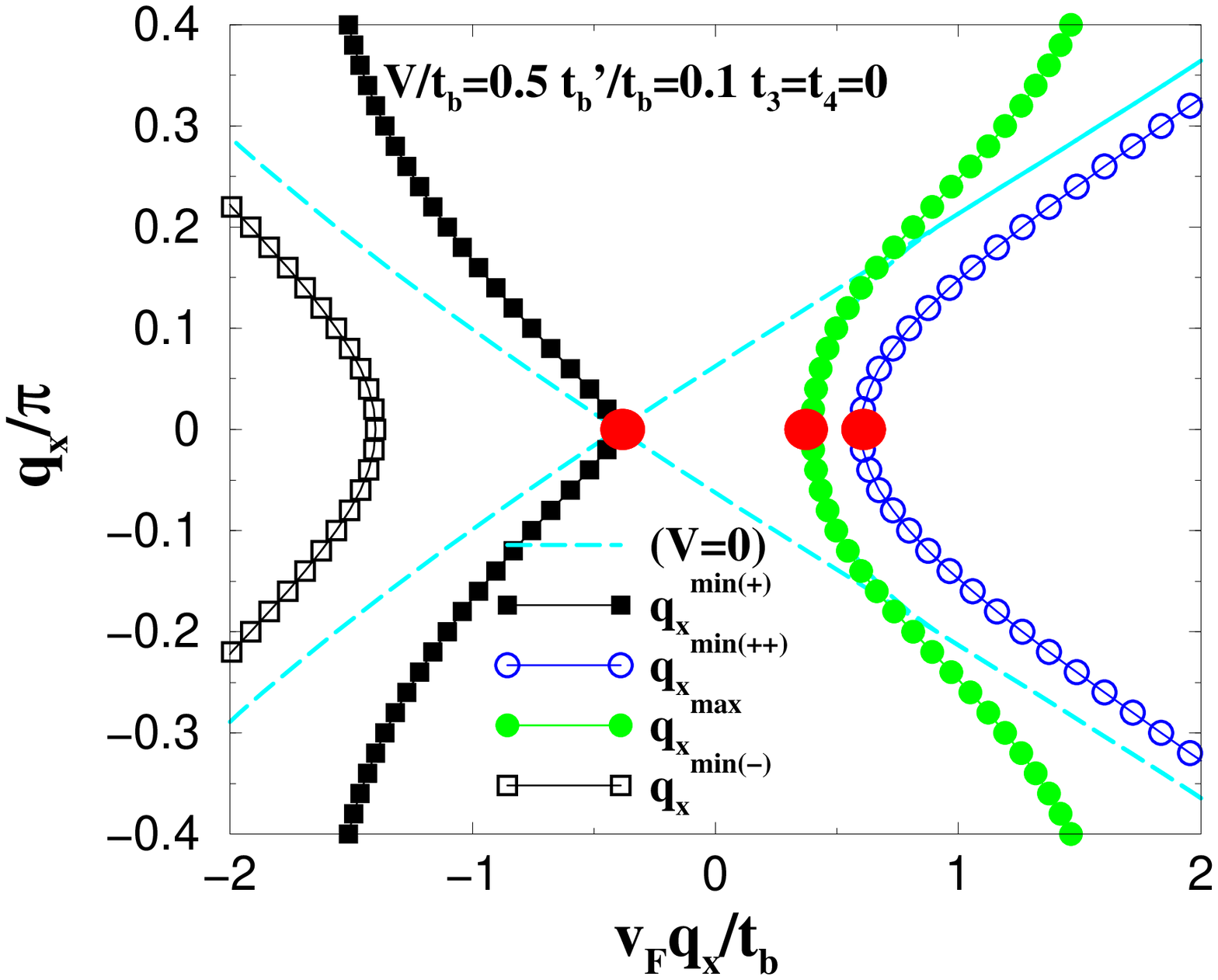}
\caption{
The same as Fig.~\ref{regionv1} for $V/t_b =0.5$.}
\label{regionv5}
\end{figure}
%
%%%%%%%%%%%%%%%%%%%%%%%%%%%%%%%%
%\begin{figure}[tbh]
%\includegraphics[width=0.38\textwidth]{t3400v1y02.eps}
%\caption{
%$q_x$ as a function of $K_y$ for $q_y/\pi=0.2$}
%\label{figv12}
%\end{figure}
%%%%%%%%%%%%%%%%%%%%%%%%%%%%%%%%%%%%%%%%%%%
%
%%%%%%%%%%%%%%%%%%%%%%%%%%%%%%%%
%\begin{figure}[tbh]
%\includegraphics[width=0.38\textwidth]{t3400v1y04.eps}
%\caption{
%$q_x$ as a function of $K_y$ for $q_y/\pi=0.4$}
%\label{figv14}
%\end{figure}
%%%%%%%%%%%%%%%%%%%%%%%%%%%%%%%%%%%%%%%%%%%
%%%%%%%%%%%%%%%%%%%%%%%%%%%%%%%
%\begin{figure}[tbh]
%\includegraphics[width=0.38\textwidth]{t3400v1y05.eps}
%\caption{
%$q_x$ as a function of $K_y$ for $q_y/\pi=0.5$}
%\label{figv15}
%\end{figure}
%%%%%%%%%%%%%%%%%%%%%%%%%%%%%%%%%%%%%%%%%%%
%%%%%%%%%%%%%%%%%%%%%%%%%%%%%
%\begin{figure}[tbh]
%\includegraphics[width=0.38\textwidth]{t3400v1qxxy.eps}
%\caption{
%.}
%\label{fig3b}
%\end{figure}
%%%%%%%%%%%%%%%%%%%%%%%%%%%%%%%%%%%%%%%%%%%
We define  
$q_{0x}$,
$q_{1x}$, $q_{2x}$, and $q_{3x}$  as 
the maximum of $q_x^{(--)}$ (at $K_y=\pm \pi/2$),
the minimum of $q_x^{(+-)}$ (at $K_y=\pm \pi/2$),
the minimum of $q_x^{(++)}$ (at $K_y=\pm \pi/2$) and
the maximum of $q_x^{(+-)}$ (at $K_y=0$ and $\pi$) 
as a function of $K_y$ when $q_y=0$
($q_{0y}=q_{1y}=q_{2y}=q_{3y}=0$), respectively
(see Fig.~\ref{figv0}), i.e.,
%${q}_{1x}$ and ${q}_{3x}$ as 
% the minimum and the maximum of
%$q_x^{(+-)}$ for $q_y=0$ as a function of $k_y$ (equivalently 
%as a function of $K_y=k_y+q_y$), 
%respectively. Since  for $q_y=0$ $q_x^{(+-)}$ (Eq.~(\ref{eqqxpmq0})) is equal to   $q_x$ (eq.~\ref{eqqx})),
%$\mathbf{q}_1$ and $\mathbf{q}_3$ do not depend on $V$,
%\begin{align}
% \mathbf{q}_1 &= (\frac{1}{\hbar k_F}(-4 t_b' + 4 t_4), 0 )\\
% \mathbf{q}_3 &= (\frac{1}{\hbar k_F}( 4 t_b' + 4 t_4), 0 )\\
%\end{align}
%For $q_y=0$,  $q_x^{+-}(K_y)$ is same as $q_x^{-+}(K_y)$.
%As $q_y$ changes, 
%As seen in Fig.~\ref{figv0},
% we obtain $q_{1x}$, $q_{2x}$, and $q_{3x}$  as 
%the minimum of $q_x^{+-}(K_y)$,
%the minimum of $q_x^{++}(K_y)$ and
%the maximum of $q_x^{+-}(K_y)$ at $q_y=0$, respectively, i.e.,
\begin{align}
 \mathbf{q}_{0} %&= -\frac{1}{\hbar v_F}[t_{\perp}(\pi/2) +t_{\perp}(3\pi/2)] \nonumber \\
%%% &=\frac{1}{\hbar v_F}
 &=\left( \frac{1}{v_F}
( -4 t_b' + 4 t_4 - 2 V),0 \right) ,
 \\
 \mathbf{q}_{1} 
%&= -\frac{1}{\hbar v_F}[t_{\perp}(\pi/2) +t_{\perp}(3\pi/2)] \nonumber \\
%%% &=\frac{1}{\hbar v_F}
 &=\left( \frac{1}{v_F}
( -4 t_b' + 4 t_4), 0 \right) ,
 \\
%\end{align}
 \mathbf{q}_{2} %&= -\frac{1}{\hbar v_F}[t_{\perp}(\pi/2) +t_{\perp}(3\pi/2)] 
%\nonumber \\
%& \ \ + \frac{1}{\hbar v_F} \sqrt{(t_{\perp}(\pi/2) -t_{\perp}(3\pi/2))^2+4 V^2}
%\nonumber \\
%%% &=\frac{1}{\hbar v_F}
 &=\left( \frac{1}{v_F}
( -4 t_b' +4 t_4 +2 V), 0 \right) ,
 \\
%\begin{align}
 \mathbf{q}_{3}
% &= -\frac{1}{\hbar v_F}[t_{\perp}(0) +t_{\perp}(\pi)] \nonumber \\
%%% &= \frac{1}{\hbar v_F}
 &= \left( \frac{1}{v_F}
(4 t_b' + 4 t_4), 0\right) .
\end{align}
%\begin{align}%
%
%\end{align}
%
When $q_{y}$ is given,
the maximums and minimums of $q_{x}^{(+-)}$
are obtained 
as a function of $K_y$
 as shown by the filled green circles and 
the filled squares 
 in Figs.~\ref{figv0},
\ref{figv11},  \ref{figv13} and \ref{figv50}.
We define $q_x^{max}(q_y)$ and $q_x^{min(+)} (q_y)$  by 
the maximums (filled green circles) 
and minimums (filled squares) 
of $q_{x}^{(+-)}$ for each $q_y$, respectively.
We also define $q_x^{min(-)}(q_y)$  by
the
value of $q_x^{(--)}$ at $K_y=\pm \pi/2$ (open black squares) and
$q_x^{min(++)}(q_y)$ by the value of $q_x^{(++)}$ at $K_y=\pm \pi /2$
(open circles).   
In Figs.~\ref{regionv1}, \ref{regionv3}, \ref{regionv4},
and \ref{regionv5}
we plot 
 $q_x^{max}(q_y)$, $q_x^{min(+)} (q_y)$,  $q_x^{min(-)}(q_y)$, 
and $q_x^{min(++)}(q_y)$ in the plane of $q_x$ and $q_y$ for
$V/t_b=0.1$, $0.3$, $0.4$ and $0.5$.
As $V$ becomes  zero, 
 $q_x^{min(++)}(q_y)$,
$q_x^{max}(q_y)$, and
$q_x^{min(-)}(q_y)$
approach to
$q_x^{min(+)}(q_y)$, $q_x^{max}(q_y)$ 
and $q_x^{min(-)}(q_y)$ at $V=0$, respectively
(cf. Fig.~\ref{figpd}).
On the other hand $q_x^{min(+)}(q_y)$
has no partner at $V=0$, 
since the filled squares in Figs.~\ref{figv0},
\ref{figv11},  \ref{figv13} and \ref{figv50}
become not the minimum but just the crossing points due to 
the folding in $K_y$ as $V$ becomes zero.
  We define $\mathbf{q}_4$ as the crossing points of 
$q_x^{min(++)}(q_y)$ and $q_x^{max}(q_y)$,
which is the extension of that in $V=0$.

%%%%%%%%%%%%%%%%%%%%%%%%%%%%%%%%%%%%%%%%%%%%%%%%%%%%%%%%%%%%%%%%%%%%%%%%%%
\begin{figure}[tbh]
\includegraphics[width=0.35\textwidth]{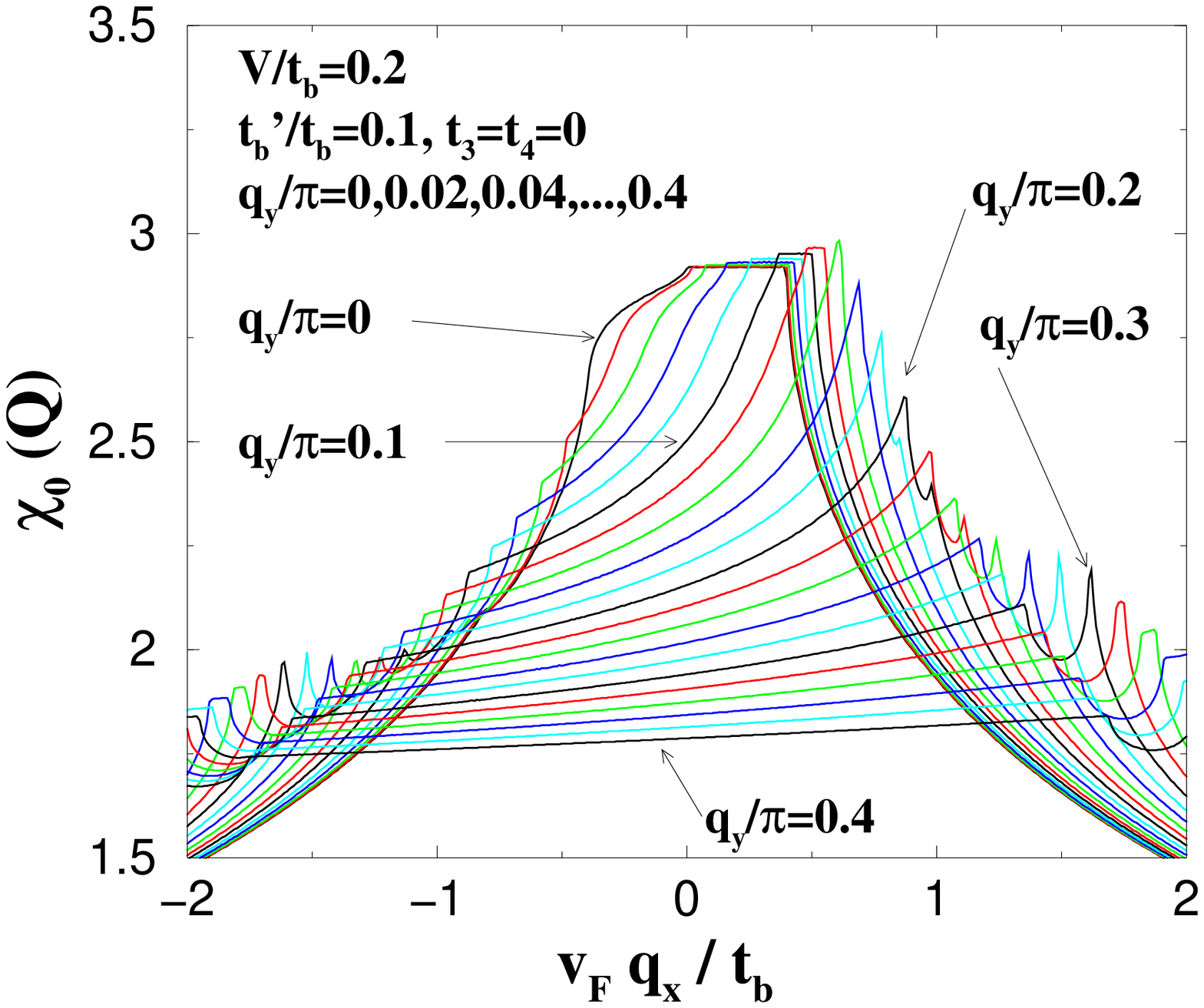}
\includegraphics[width=0.35\textwidth]{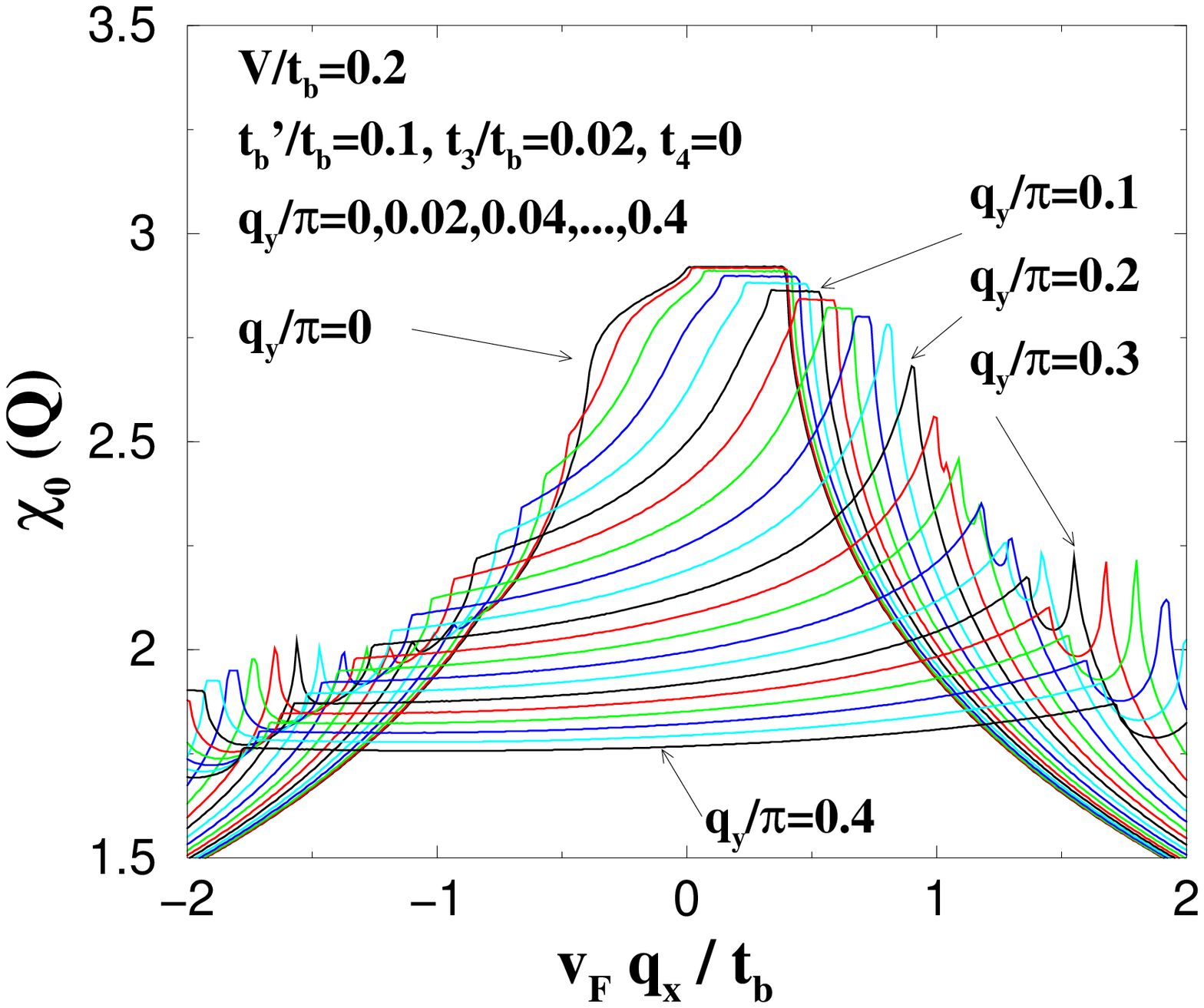}
\includegraphics[width=0.35\textwidth]{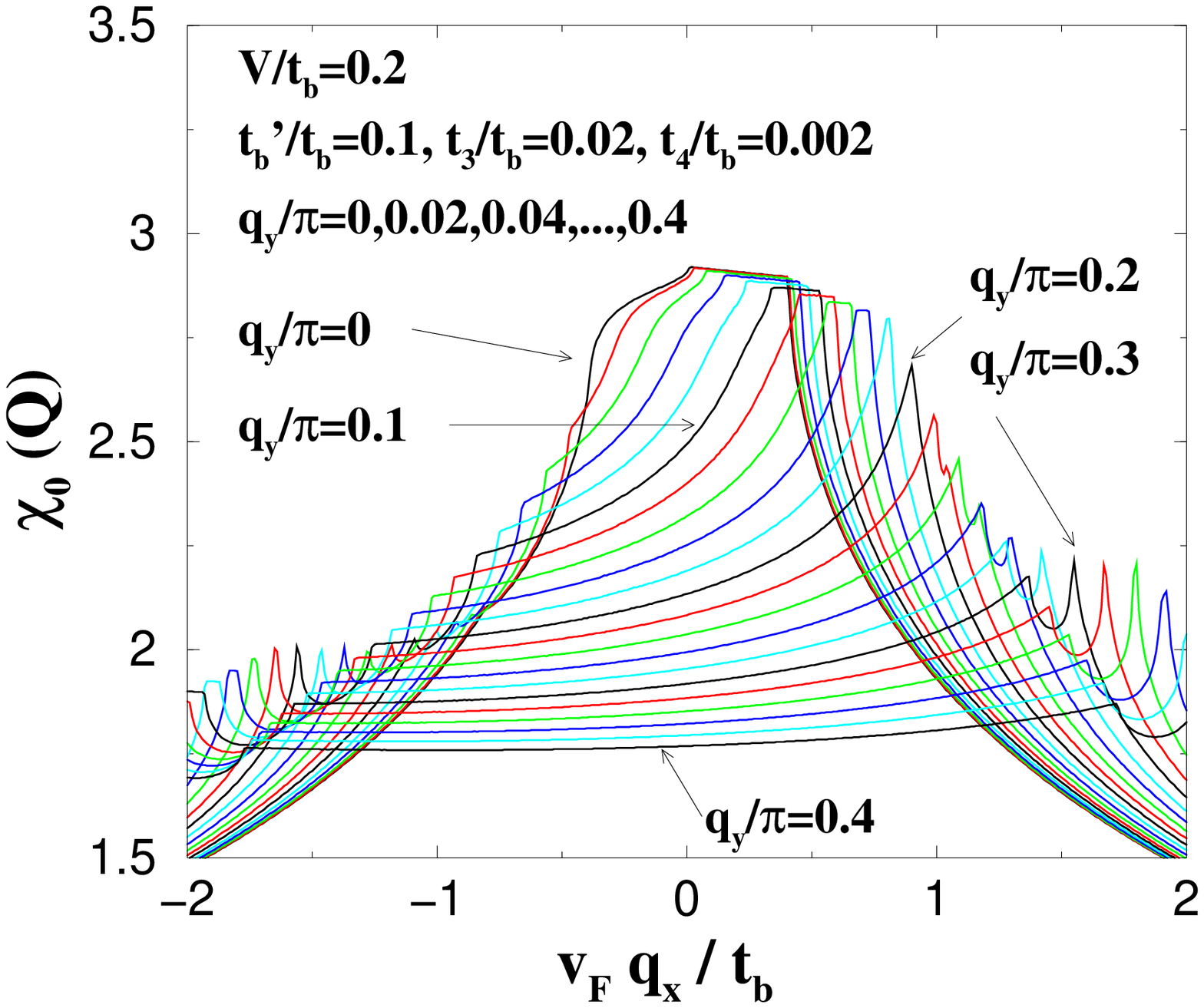}
\caption{$\chi_0(\mathbf{Q})$ at $T=0$  
as a function of $q_x$. The parameters are the same as in
Fig.~\ref{chiqt0} but $V/t_b=0.2$. 
}
\label{figchiv02}
\end{figure}
%%%%%%%%%%%%%%%%%%%%%%%%%%%%
%%%%%%%%%%%%%%%%%%%%%%%%%%%%%%%%%%%%%%%%%%%%%%%%%%%%%%%%%%%%%%%%%%%%%%%%%%%
\begin{figure}[tbh]
\includegraphics[width=0.35\textwidth]{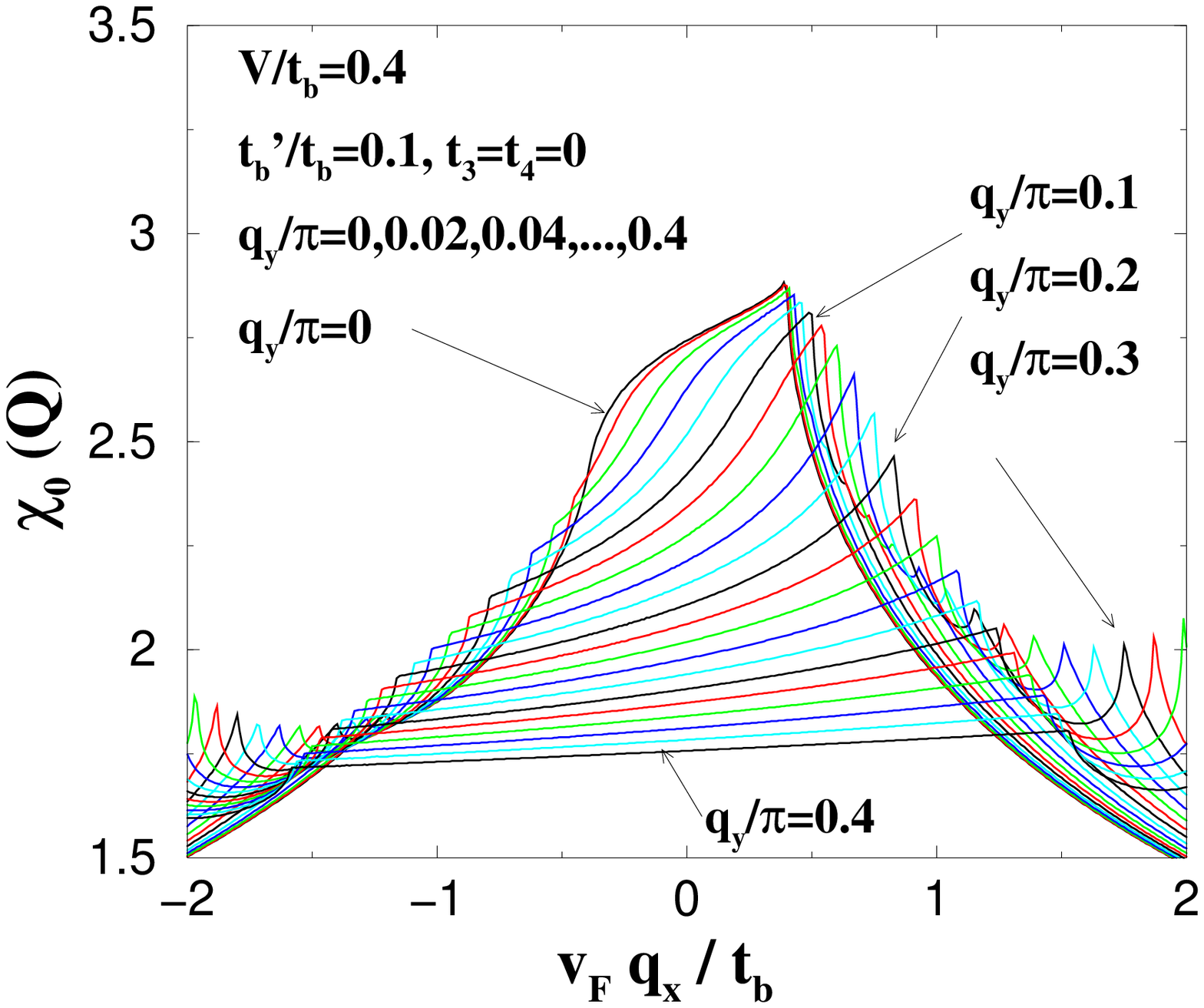}
\includegraphics[width=0.35\textwidth]{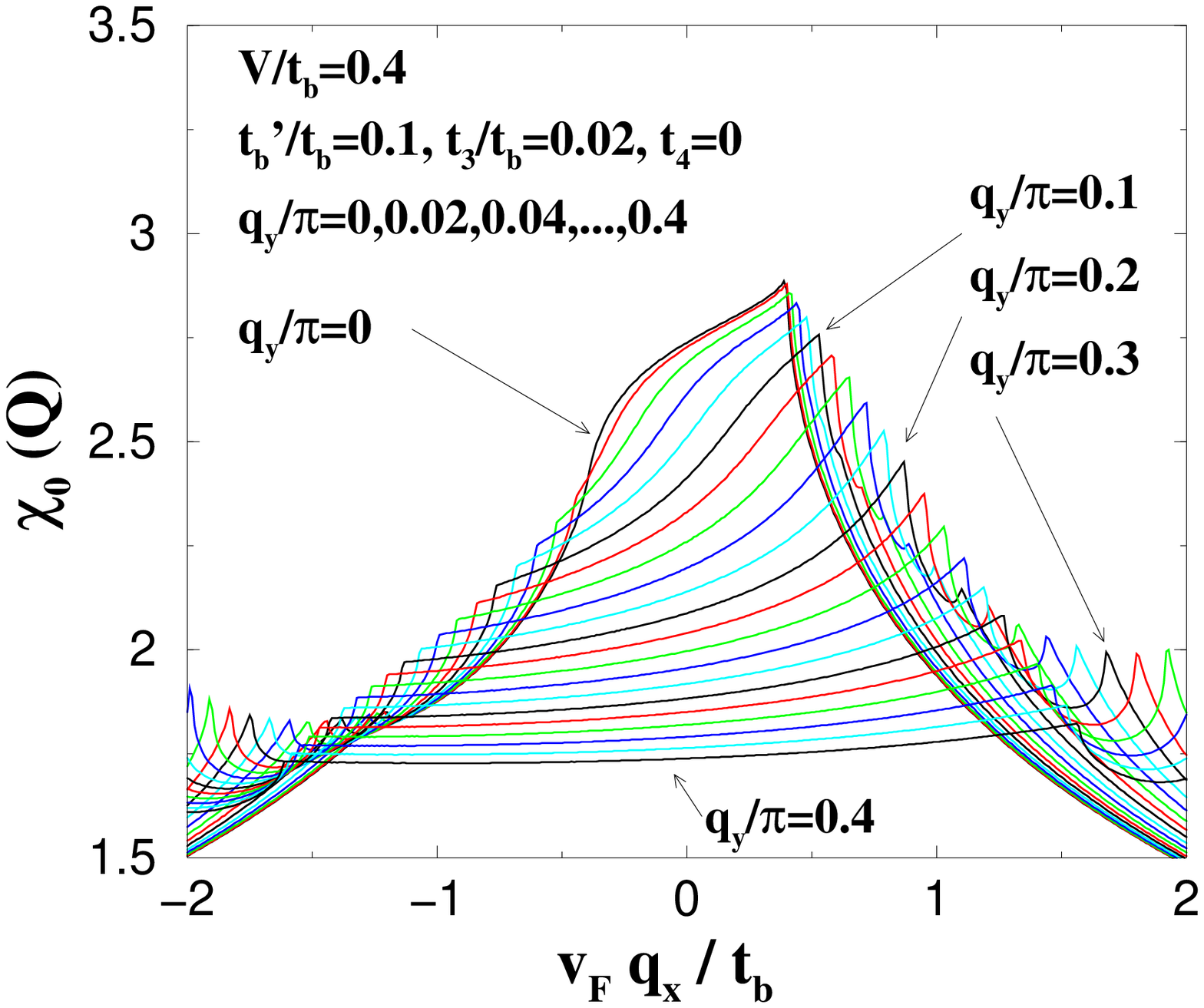}
\includegraphics[width=0.35\textwidth]{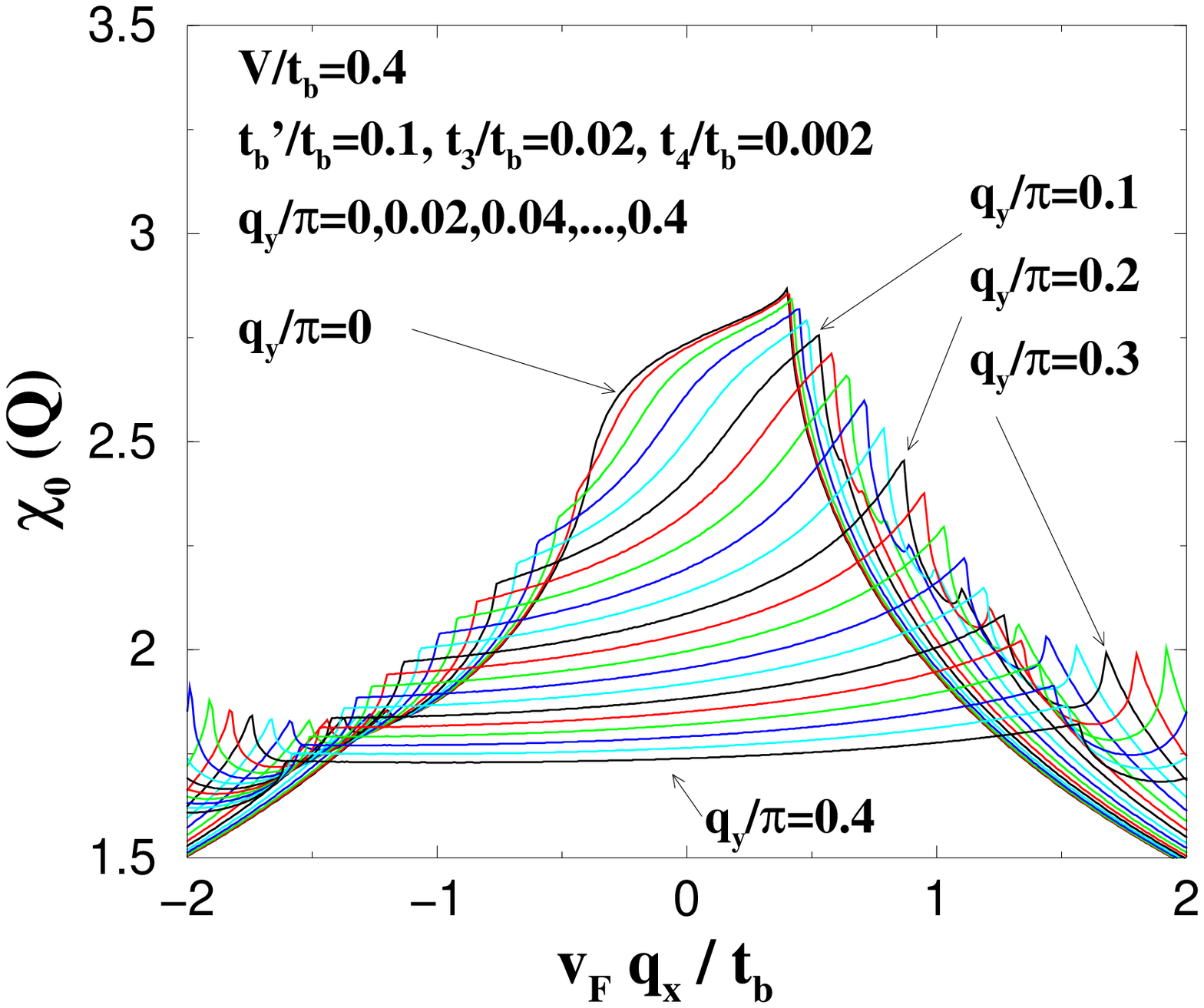}
\caption{The same as Fig.~\ref{figchiv02} with $V/t_b=0.4$.
}
\label{figchiv04}
\end{figure}
\begin{figure}[tbh]
\includegraphics[width=0.38\textwidth]{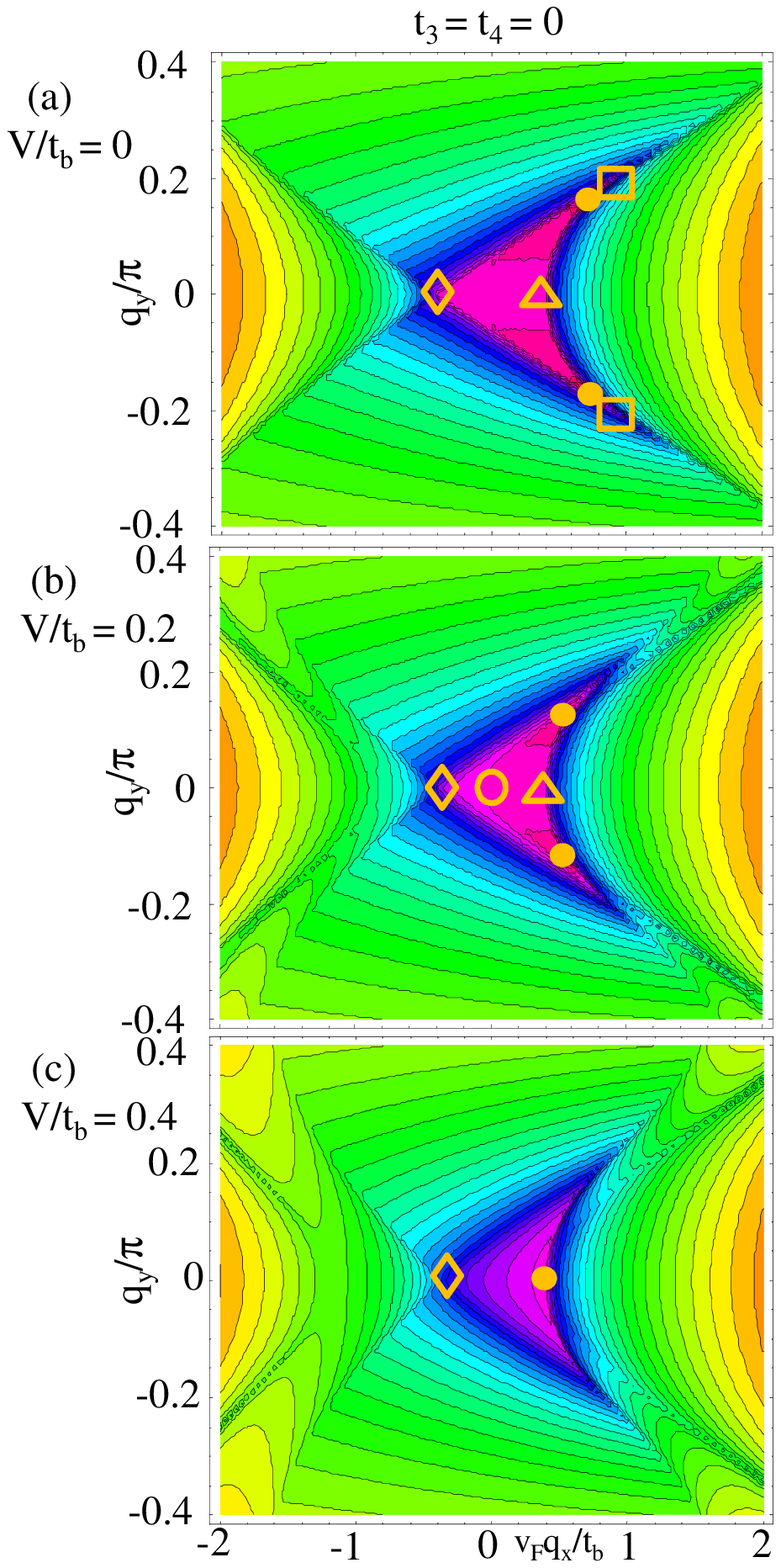}
\caption{%vcon.eps
The contour plot of $\chi_0(\mathbf{Q}_0+\mathbf{q})$.
The filled circles show the location of the maximum
(best nesting vector).
The diamonds, the open circles, the triangles, and the squares are
$\mathbf{q}_1$, $\mathbf{q}_2$, $\mathbf{q}_3$, and $\mathbf{q}_4$,
respectively. 
We take $t_b'/t_b=0.1$, $t_3=t_4=0$.
}
\label{vcon}
\end{figure}
\begin{figure}[tbh]
\includegraphics[width=0.38\textwidth]{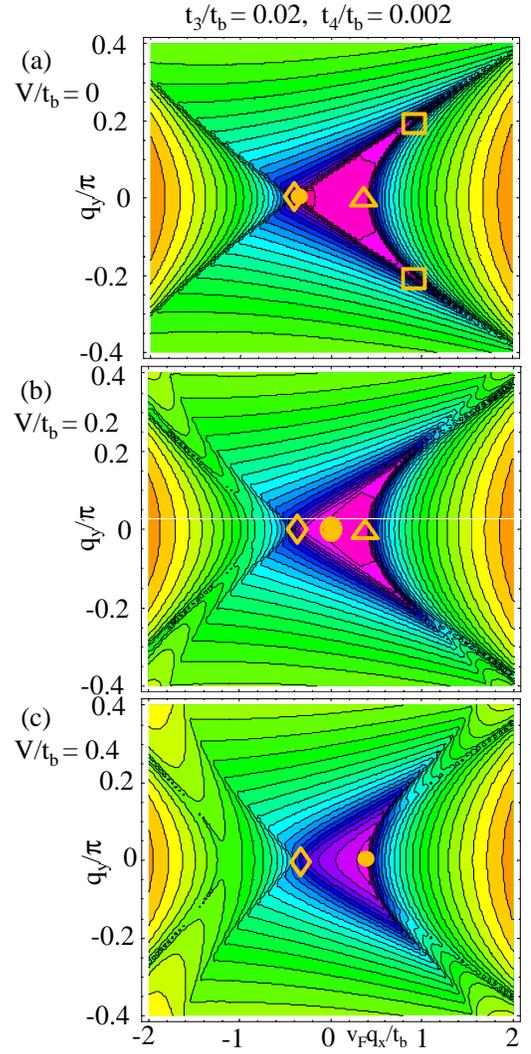}
\caption{The same as Fig.~\ref{vcon} with
$t_b'/t_b=0.1$, $t_3/t_b=0.02$ and $t_4/t_b=0.002$
}
\label{vcon2}
\end{figure}
\begin{figure}[tbh]
\includegraphics[width=0.38\textwidth]{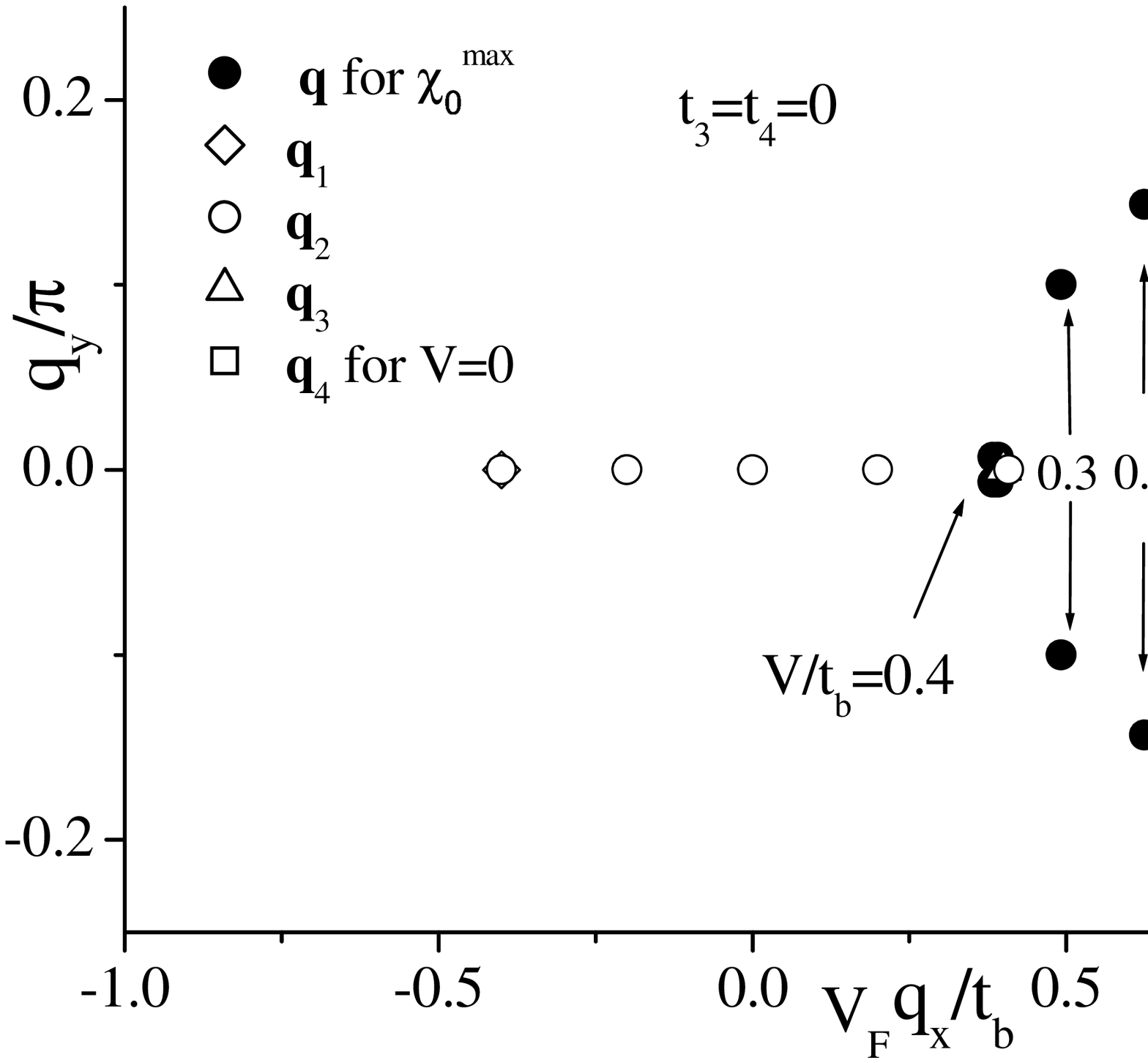}
\caption{
Open diamond, open triangle and open squares are
$\mathbf{q}_1$, $\mathbf{q}_3$ and $\mathbf{q}_4$ for $V=0$,
respectively.
Open circles and closed circles are $\mathbf{q}_2$ and
the locations of the maximum of $\chi_0(\mathbf{Q})$
(best nesting vector), respectively,
 for $V/t_b=0$, $0.1$, $0.2$, $0.3$ and $0.4$.
We take $t_b'/t_b=0.1$, $t_3=t_4=0$.
}
\label{figqxqy}
\end{figure}

%%%%%%%%%%%%%%%%%%%%%%%%%%%%
We plot $\chi_0(\mathbf{Q}_0+\mathbf{q})$ as a function of $q_x$
for several values of $q_y$ in 
 Fig.~\ref{figchiv02} ($V/t_b=0.2$) and Fig.~\ref{figchiv04} ($V/t_b=0.4$)
for the parameters $t_b'/t_b=0.1$ and some values of $t_3$ and $t_4$. 
The contour plots of the $\chi_0(\mathbf{Q}_0+\mathbf{q})$ in the $k_x - k_y$
plane are shown 
in Fig.~\ref{vcon} ($t_3=t_4=0$) and Fig.~\ref{vcon2} 
($t_3/t_b=0.02$, $t_4/t_b=0.002$)
for $V/t_b=0$, $0.2$ ($V/t_b'=2$) and $0.4$ ($V/t_b'=4$).
When $0 < V < 4 t_b'$, $q_{1x}<q_{2x}<q_{3x}$. In this case 
$\chi_0(\mathbf{Q}_0+ \mathbf{q})$ has a plateau-like maximum 
in the ``\textit{sweptback}'' region enclosed by $\mathbf{q}_2$, 
$\mathbf{q}_4$, and $\mathbf{q}_3$,
 as shown in Figs.~\ref{regionv1} and \ref{regionv3}.
This region shrinks to the point $\mathbf{q}_3$ when
$V/t_b'=4$ as shown in Fig.~\ref{regionv4}.
The absolute maximum of $\chi_0(\mathbf{Q}_0+\mathbf{q})$
 occurs near $\mathbf{q}_4$ if $t_3=t_4=0$.
The effects of $t_3$ and $t_4$ on $\chi_0(\mathbf{Q}_0+\mathbf{q})$
are the same as these at $V=0$;
A finite $t_3$ suppresses $\chi_0(\mathbf{Q}_0+\mathbf{q})$ at $q_y \neq 0$
and $t_4$ lifts the degeneracy at $q_{2x} \leq q_x \leq q_{4x}$.
If $ V > 4 t_b'$,  we obtain $q_{1x}<q_{3x}<q_{2x}$ 
and there are no region where $\chi_0(\mathbf{Q})$
has a plateau-like maximum as shown 
in Figs.~\ref{regionv4}, \ref{regionv5} and \ref{figchiv04}. 
In that case the effects of $t_3$ and $t_4$ are small.
In Fig.~\ref{figqxqy}, $\mathbf{q}_2$ and 
$\mathbf{q}$ which gives the maximum of $\chi_0(\mathbf{Q}_0+\mathbf{q})$
(i.e., the best nesting vector)
are shown for some values of $V/t_b$ in the case of $t_3=t_4=0$. 
The best nesting vector moves to $\mathbf{q}_3$ as $V/t_b$ approaches to $0.4$.
%%%%%%%%%%%%%%%%%%%%%%%%%%%%
%%%%%%%%%%%%%%%%%%%%%%%%%%%%%%%%%%%%%%%%%%%%%%%%%%%%%%%%%%%%%%%%%%%%%%%%%%%
%%%%%%%%%%%%%%%%%%%%%%%%%%%%
%%%%%%%%%%%%%%%%%%%%%%%%%%%%%%%%%%%%%%%%%%%

The negative Hall constant in some region 
of the magnetic field\cite{ribault,matsunaga2} has been 
explained by the $t_3$ and $t_4$ terms\cite{zanchi96}.
When $V=0$, the terms with $t_3/t_b=0.02$ and $t_4/t_b=0.002$  
make the absolute maximum of $\chi_0(\mathbf{Q})$
in the zero magnetic field to be at $\mathbf{q}_1$ (best nesting vector),
while the best nesting vector is located near $\mathbf{q}_4$ if $t_3=t_4=0$,
as shown in Fig.~\ref{chiqt0}.
The negative Hall constant is possible, since $q_{1x} <0$. 
If $V /t_b' >0$ and $t_3$ and $t_4$ are the same as
above, the best nesting vector is $\mathbf{q}_2$ 
(see the lower figures
in Fig.~\ref{figchiv02} and the middle figure in Fig.~\ref{vcon2}). 
As far as $V  <2 t_b'-2 t_4$, the negative Hall constant is possible
since $q_{2x} <0$. If $V  >2 t_b'-2 t_4$, however,
the best nesting vector $\mathbf{q}_2$ has the positive $x$ component, 
as seen in the lower figures in
Fig.~\ref{figchiv02} and Fig.~\ref{figchiv04}.
Therefore, the negative Hall constant is difficult to be stabilized when 
$V >2 t_b' - 2 t_4$.
Recently, the authors\cite{kishigi2007}
 have numerically obtained  the phase diagram for the quantum Hall
effect as a function of the magnetic field and periodic potential $V$.
We have shown that the negative Hall constant  ($N=-2$) appears only 
in the region $ 0.03 \lesssim V/t_b \lesssim 0.2$
($0.3 \lesssim V/t_b' \lesssim 2$)
for the parameters $t_b'/t_b=0.1$, $t_3/t_b=0.02$ and $t_4/t_b=0.002$
(the upper figure of Fig.~12 in Ref. [\onlinecite{kishigi2007}]).
 That result can be understood by the fact 
that for $V>2 t_b'-2 t_4$ the best nesting
vector has the positive $x$ component.
The existence of the negative Hall constant for  $V/t_b' \gtrsim 0.3$
is understood by the effect of
  $V$ that will make  $\chi_0(\mathbf{Q}_0+\mathbf{q})$ 
at $\mathbf{q} \approx \mathbf{q}_4$ to be smaller. 
Experimentally, a negative Hall effect  is observed 
when the system is cooled  slowly  (less than $0.03$K/s) 
and the external magnetic field region for the negative Hall effect
becomes larger as the cooling rate becomes slower  (the slowest 
cooling rate is $0.000 09$ K/s).\cite{matsunaga2} 
It is expected that the
magnitude of the periodic potential $V$ becomes 
larger at the slower colling rate.
Therefore, we can conclude from the existence of the negative Hall effect
in (TMTSF)$_2$ClO$_4$
 that $V < 2 t_b'-2 t_4$. 
The value of 
$V$ estimated from the magnetic-field-angle dependence of the 
conductivity\cite{yoshino2003,lebed2005,ha2006}
is close to the border of this condition.
\clearpage
\section{summary and discussions}
We have studied the nesting vector and $\chi_0(\mathbf{Q})$
in the quasi-one dimensional systems having the imperfectly nested Fermi
surface (the imperfectness is measured by $t_b'$). 
We have obtained the plateau-like maximum of 
$\chi_0(\mathbf{Q})$ when $\mathbf{Q}$ is in the
\textit{sweptback} region  with the apexes $\mathbf{q}_1$ and $\mathbf{q}_4$. 
% of \textit{the tip of the arrow}. 
The absolute maximum of
$\chi_0(\mathbf{Q})$ is obtained near but not at $\mathbf{Q} = 
\mathbf{Q}_0+\mathbf{q}_4$ if $t_3=t_4=0$.
When the periodic potential $V$ is finite but not as large as $4 t_b'$
(which is thought to be the case in (TMTSF)$_2$ClO$_4$), 
the ``sweptback'' region (with apexes $\mathbf{q}_2$ and $\mathbf{q}_4$)
 becomes smaller
as $V$ increases and shrinks to $\mathbf{q}_3$ 
when $V=4 t_b'$. 
The best nesting vector moves to $\mathbf{Q} 
\approx 
\mathbf{Q}_0+\mathbf{q}_3$. The absolute maximum of $\chi_0(\mathbf{Q})$
is located at $\mathbf{Q}  = \mathbf{Q}_0+\mathbf{q}_3$
when $V> 4 t_b'$.
The negative Hall coefficient observed 
in the field-induced spin density wave states
in some region of the magnetic field is shown to be possible only when
$V < 2 t_b'-2 t_4$, in which case the vectors $\mathbf{q}$'s giving the  
plateau-like maximum of $\chi_0(\mathbf{Q}_0+\mathbf{q})$ 
(``\textit{sweptback}'' region)
can have the negative $x$ component,  ($q_{2x} <0$).
Therefore, we conclude that $V$ should be smaller than $2t_b'-2t_4$
in (TMTSF)$_2$ClO$_4$, where the sign reversal of the Hall effect 
has been observed.

Recently, a lot of interest is attracted by the 
quasi-one-dimensional conductor (Per)$_{2}$ \textit{M}(mnt)$_2$ 
(where Per = perylene, mnt = maleonitriledithiolate and 
\textit{M} = 
Au and Pt)\cite{graf2004,canadell2004,mcdonald2005,lebed2007,graf2007}.
The charge density wave (CDW) state is realized 
in (Per)$_{2}$\textit{M}(mnt)$_2$, and
the successive transitions of the field-induced CDW has been observed 
in high magnetic field\cite{graf2004}
in contrast to the field-induced SDW in (TMTSF)$_2$ClO$_4$.
This material has a similar band structure as (TMTSF)$_2$ClO$_4$,
but
the origin of the pairs of the quasi-one-dimensional Fermi surface
in (Per)$_{2}$\textit{M}(mnt)$_2$  is
different from that in (TMTSF)$_2$ClO$_4$. 
The origin of the 
four pairs of the quasi-one-dimensional Fermi surface 
in (Per)$_{2}$\textit{M}(mnt)$_2$ is the 
existence of four perylene molecules in the unit cell in the perpendicular plane 
to the conduction axis\cite{canadell2004},
while 
the origin of the 
two pairs of the quasi-one-dimensional Fermi surface  in (TMTSF)$_2$ClO$_4$ 
is the periodic potential caused by the anion ordering.
It will be interesting to study the similarity and the difference between two 
materials, 
since the spin susceptibility $\chi_0(\mathbf{Q})$ and the charge susceptibility 
$\chi_c(\mathbf{Q})$
for the non-interacting system  have the same $\mathbf{Q}$ 
dependence caused by the nesting properties
of the Fermi surface, 
except for the effects of the Zeemen splitting of the Fermi surface, 
which play
important role only for CDW.

%%%%%%%%%%%%%%%%%%%%%%%%%%%%%%%%%%%%%%%%%%%
\begin{acknowledgments}
%\section{Acknowledgment}
This work is partly supported by a Grant-in-Aid for the Promotion 
of Science and Scientific Research on Priority 
Areas (Grant No. 18028021) from the 
Ministry of Education, Culture, Sports, Science and Technology, Japan. 
\end{acknowledgments}
%\clearpage
%\newpage
%%%%%%%%%%%%%%%%%%%%%%%%%%

\end{document}